\RequirePackage{etoolbox}
\csdef{input@path}{%
 {sty/}
 {img/}
}%
\csgdef{bibdir}{bib/}

\documentclass[ba]{imsart}
\usepackage{amsthm}
\usepackage{amsmath}
\usepackage{natbib}
\usepackage[colorlinks,citecolor=blue,urlcolor=blue,filecolor=blue,backref=page]{hyperref}
\usepackage{graphicx}
\usepackage{amsthm}
\usepackage{amsmath}
\usepackage{natbib}
\usepackage[colorlinks,citecolor=blue,urlcolor=blue,filecolor=blue,backref=page]{hyperref}
\usepackage{graphicx}
\RequirePackage[OT1]{fontenc}
\RequirePackage{amsthm,amsmath}
\RequirePackage{natbib}
\usepackage[utf8]{inputenc}
\usepackage{natbib}
\usepackage{amsthm}
\usepackage{amsfonts}
\usepackage{mathtools}
\usepackage{amssymb}
\usepackage{graphicx}
\usepackage{float}
\usepackage{pgfplotstable}
\usepackage{booktabs}
\usepackage{colortbl}
\usepackage{multirow}
\usepackage{array}
\usepackage{ragged2e}
\usepackage{longtable}
\usepackage{tabu}
\usepackage{rotating}
\usepackage{lscape}
\usepackage{afterpage}
\usepackage{multirow}
\usepackage{subfigure}
\usepackage{listings}
\usepackage{color}
\usepackage{dsfont}
\usepackage{pdflscape}
\usepackage[left=4.5cm,right=4.5cm,top=2.5cm,bottom=2.5cm,footnotesep=0.5cm]{geometry}
\usepackage{tabularx} 
\newcolumntype{K}[1]{>{\centering\arraybackslash}p{#1}}
\newcolumntype{Q}[1]{>{\columncolor[gray]{0.8}\centering\arraybackslash}p{#1}}

\newcommand\norm[1]{\left\lVert#1\right\rVert}

\startlocaldefs
\endlocaldefs

\begin{document}

\begin{frontmatter}
\title{Regret-based Selection for \\Sparse Dynamic Portfolios}
\runtitle{Regret-based Selection}

\begin{aug}
\author{\fnms{David} \snm{Puelz}\thanksref{addr1}\ead[label=e1]{david.puelz@utexas.edu}\ead[label=u1,url]{http://faculty.mccombs.utexas.edu/David.Puelz/}},
\author{\fnms{P. Richard} \snm{Hahn}\thanksref{addr2}\ead[label=e2]{richard.hahn@chicagobooth.edu}}
\and
\author{\fnms{Carlos M.} \snm{Carvalho}\thanksref{addr1}
\ead[label=e3]{carlos.carvalho@mccombs.utexas.edu}}

\runauthor{Puelz, Hahn and Carvalho}

\address[addr1]{
	University of Texas McCombs School of Business
}

\address[addr2]{
University of Chicago Booth School of Business
}

\end{aug}

\begin{abstract}
This paper considers portfolio construction in a dynamic setting.  We specify a loss function comprised of utility and complexity components with an unknown tradeoff parameter. We develop a novel regret-based criterion for selecting the tradeoff parameter to construct optimal sparse portfolios over time.


\end{abstract}

%

\end{frontmatter}

\section{Introduction}

Practical investing requires balancing portfolio optimality and simplicity. In other words, investors desire well-performing portfolios that are \textit{easy} to manage, and this preference is driven by many factors. Managing large asset positions and transacting frequently is expensive and time-consuming, and these complications arise from both trading costs and number of assets available for investment.  For the individual investor, these challenges are strikingly amplified.  Their choice set for investment opportunities is massive and includes exchange-traded funds (ETFs), mutual funds, and thousands of individual stocks. This raises the question: \textit{How does one invest optimally while keeping the simplicity (sparsity) of a portfolio in mind?} Further challenges arise when sparse portfolio selection is placed in a dynamic setting. The investor will want to update her portfolio over time as future asset returns are realized while maintaining her desire for simplicity.  Institutions offering retirement accounts to employees represents a practical example of this problem (i.e., a 403(b) plan at public universities). An employee must choose from a large menu of funds to invest in her retirement account.  This paper takes the perspective of an individual investor who is faced with the task of selecting funds and constructing portfolios over time.

The focus of this paper are loss functions that balance utility and sparsity myopically for each time $t$:
\begin{equation} \label{overalllossfirst}
	\begin{split}
		\textbf{L}_{\lambda_t}(w_t,\tilde{R}_t) = \mathcal{L}(w_t,\tilde{R}_t) + \Phi(\lambda_t,w_t),
	\end{split}
\end{equation}where $\tilde{R}_t$ is a vector of future asset returns, $\mathcal{L}$ is the negative utility of an investor, $w_t$ is a vector of portfolio weights, $\Phi$ is a function that encourages sparsity in $w_t$, and $\lambda_t$ is a penalty parameter governing the degree to which the complexity function $\Phi$ is influential in the overall loss function.  Special attention must be paid to $\lambda_t$, the parameter that governs the utility-sparsity tradeoff.  If it is known a priori, the investor's optimal portfolio may be found for each time by routine minimization of the expectation of (\ref{overalllossfirst}).  By contrast, this paper considers the more challenging case where this parameter is unknown and may be thought of as part of the investor's decision. 

The interplay between dynamics, utility and portfolio simplicity in the investor's portfolio decision are viewed through the lens of regret. Assuming the existence of a desirable target portfolio, we define regret as the difference in loss (or negative utility) between the simple portfolio and the target; our investor would like to hold a simple portfolio that is ``almost as good as" a target. This paper distills a potentially intractable dynamic selection procedure into one which requires specification of only a single threshold of regret.  At the outset, the investor need only answer the question: \textit{With what degree of certainty do I want my simple portfolio to be no worse than the target portfolio?}  Put differently: \textit{What maximum probability of regretting my portfolio decision am I comfortable with?}


Once the regret threshold is specified, the investor's preference for portfolio simplicity will automatically adjust over time to accommodate this threshold.  In other words, the penalty parameter $\lambda_t$ continuously adjusts to satisfy the investor's regret tolerance.  In one period, her portfolio may only need to be invested in a small number of assets to satisfy her regret threshold.  However, that same portfolio may be far off from the target in the next period, requiring her to invest in more assets to accommodate the same level of regret.  This thought experiment illustrates that our procedure, although requiring a static regret tolerance to be specified at the outset, results in investor preferences for sparsity that are \textit{dynamic}.

The regret-based selection approach presented in this paper is related to the \textit{decoupling shrinkage and selection (DSS)} procedure from \cite{HahnCarvalho} and \cite{puelz2016variable}.  In both papers, loss functions with explicit penalties for sparsity are used to summarize complex posterior distributions.  Posterior uncertainty is then used as a guide for variable selection in static settings.  This paper expands on these notions by developing a regret-based metric for selection and placing it within a dynamic framework. The important innovations presented herein are (\textit{i}) The use of investor's \textit{regret} for selection and (\textit{ii}) The development of a principled way to choose a \textit{dynamic} penalty parameter $\lambda_t$ (and thus select a portfolio) for a time-varying investment problem.

A key finding of this paper is that sparse portfolios and their more complex (or dense) counterparts are often very similar from an \textit{ex ante} regret perspective. More surprisingly, this similarity often persists \textit{ex post}.  This gives credence to a common piece of investing advice: ``Don't worry about investing in a variety of funds; just buy the market."

\subsection{Previous research}

The seminal work of \cite{markowitz1952portfolio} provides the foundation of utility design for portfolio optimization related to this paper. One area of relevant research highlighting Bayesian approaches to this problem may be found in \cite{zhou2014bayesian}.  In this paper, the authors consider portfolio construction with sparse dynamic latent factor models for asset returns. They show that dynamic sparsity improves forecasting and portfolio performance.  However, sparsity in their context is induced at the factor loading level, not the portfolio decision.  In contrast, our methodology seeks to sparsify the portfolio decision directly for \textit{any} generic dynamic model. 

Additional areas of research focus on the portfolio selection problem, particularly in stock investing and index tracking.  \cite{polson1999bayesian} consider the S\&P 500 index and develop a Bayesian approach for large-scale stock selection and portfolio optimization from the index's constituents.  Other insightful Bayesian approaches to optimal portfolio choice include \cite{johannes2014sequential}, \cite{irie2016bayesian}, \cite{zhao2016dynamic}, \cite{gron2012optimal}, \cite{jacquier2010simulation}, \cite{puelz2015optimal} and \cite{pettenuzzo2015optimal}.  Methodological papers exploring high dimensional dynamic models relevant to this work are \cite{carvalho2007dynamic} and \cite{wang2011dynamic}.

\section{Overview}

The focus will be loss functions of the following form:
\begin{equation} \label{overallloss}
	\begin{split}
		\textbf{L}_{\lambda_t}(w_t,\tilde{R}_t) = \mathcal{L}(w_t,\tilde{R}_t) + \Phi(\lambda_t,w_t)
	\end{split}
\end{equation}where $\mathcal{L}$ is the negative utility of an investor, $\tilde{R}_t$ is a vector of $N$ future asset returns, $w_t$ is a vector of portfolio weights, and $\lambda_t$ is a penalty parameter governing the degree to which the complexity function $\Phi$ is influential in the overall loss function. Let the future asset returns be generated from a model parameterized by $\Theta_t$ so that $\tilde{R_t} \sim \Pi(\Theta_{t})$ and $\Pi$ is a general probability distribution.

The time-varying preferences in (\ref{overallloss}) take into account an investor's negative utility as well as her desire for portfolio simplicity.  Optimization of (\ref{overallloss}) in practice poses an interesting challenge since there is uncertainty in model parameters $\Theta_t$ and future asset returns $\tilde{R}_t$.  Also, the penalty parameter $\lambda_t$ is not known by the investor a priori, making her risk preferences ambiguous in portfolio complexity. A further obvious complication is that all of these unknowns are varying in time.

We propose a three-step approach to constructing a sequence of sparse dynamic portfolios.  This procedure will be based on an investor's regret from investing in an alternative (target) portfolio defined by $w_t^*$.  The three general steps are:
\begin{enumerate}
	\item \textit{Model specification:} Model the future asset returns $\tilde{R_t} \sim \Pi(\Theta_{t})$.
\vspace{3mm} 
	\item \textit{Loss specification:} Specify $\mathcal{L}$ and $\Phi$ in Loss (\ref{overallloss}).  Then, the expected loss given by $\textbf{L}_{\lambda_t}(w_t) = \mathbb{E}[\textbf{L}_{\lambda_t}(w_t,\tilde{R}_t)]$ may be minimized for a sequence of $\lambda_t \hspace{1mm} \forall t$. Define the collection of optimal portfolios in the cross-section as $\{ w_{\lambda_t}^* \}$.
\vspace{3mm} 
	\item \textit{Regret-based summarization:} Compare regret-based summaries of the optimal portfolios versus a target portfolio $w_t^*$ by thresholding quantiles of a \textit{regret} probability distribution, where regret as a random quantity is given by $\rho(w_{\lambda_t}^{*},w_t^*,\tilde{R}_t)$.  This random variable is a function of a sparse portfolio decision $w_{\lambda_t}^*$, the target portfolio $w_{t}^*$, and future asset returns.
\end{enumerate}The expectation and probability are both taken over the joint distribution of unknowns $(\tilde{R}_t, \Theta_t \mid \textbf{R}_{t-1})$ where $\textbf{R}_{t-1}$ is observed asset return data.  Flexibly, the regret function $\rho(w_{\lambda_t}^{*},w_t^*,\tilde{R}_t)$ can be any metric that is a function of the portfolio weights and unknowns and may be constructed using any target portfolio $w_t^*$.   In this paper, we consider the difference in loss between the sparse portfolio decision and the target portfolio  $\rho(w_{\lambda_t}^{*},w_t^*,\tilde{R}_t) = \mathcal{L}(w_{\lambda_t}^{*},\tilde{R}_t) - \mathcal{L}(w_t^*,\tilde{R}_t)$ as our measure of regret in keeping with the usual decision theoretic definition. 

We can now see how portfolio sparsity appears in the dynamic setting. The dynamic model given by $\tilde{R}_t \sim \Pi(\Theta_t)$ interacts with the portfolio decision $w_t$ via the expected loss minimization in step 2.  Iterating step 3 over time gives a sequence of sparse portfolio summaries $\{ w_{\lambda_t^*}^* \}$ where $\lambda_t^*$ provides an index for the selected sparse decision. Ultimately, these sparse portfolio summaries select which subsets of assets are relevant for our ``simple portfolio".  



The details of this procedure will be fleshed out in the following subsection.

\subsection{Details of Regret-based summarization} \label{details}
 In following section, we discuss the specifics of the regret-based summarization procedure.  We focus on expanding on step 3 of the methodology which represents the main innovation of this paper. Model specification and fitting as well as loss specification comprising steps 1 and 2 are only highlighted; a detailed formulation of these first two steps is presented in the Application section. 

Suppose that we have inferred a model for future asset returns given observed data $\textbf{R}_{t-1}$ that is parameterized by the vector $\Theta_t$: $\tilde{R}_t \sim \Pi(\Theta_t \mid \textbf{R}_{t-1})$. Let the resulting posterior distribution of parameters and future asset returns be $p(\Theta_t, \tilde{R}_t \mid \textbf{R}_{t-1})$.  For example, $\Pi$ may be parameterized by dynamic mean and variance parameters $\Theta_t = (\mu_t,\Sigma_t)$.  Further, suppose we have specified the utility and complexity components of the investor's loss function: $\textbf{L}_{\lambda_t}(w_t,\tilde{R}_t) = \mathcal{L}(w_t,\tilde{R}_t) + \Phi(\lambda_t,w_t)$.

For each investing period $t$, we obtain a sequence of portfolio decisions indexed by $\lambda_t$ by optimizing the expected loss function $\mathbb{E}[\textbf{L}_{\lambda_t}(w_t,\tilde{R}_t)]$.  Regret-based summarization is an approach to select the appropriate optimal decision from the collection (i.e., select $\lambda_t$) for each time $t$, and this choice may be visualized by using sparsity-regret tradeoff plots.

Revisiting the regret, samples of the $\tilde{R}_t$ margin from posterior distribution $(\tilde{R}_t, \Theta_t \mid \textbf{R}_{t-1})$ define the distribution for the regret random variable $\rho(w_{\lambda_t}^{*},w_t^*,\tilde{R}_t)$ given by a difference in loss:
\begin{equation} \label{regretdefn}
	\begin{split}
		\rho(w_{\lambda_t}^{*},w_t^*,\tilde{R}_t) &= \mathcal{L}(w_{\lambda_t}^{*},\tilde{R}_t) - \mathcal{L}(w_t^*,\tilde{R}_t),
	\end{split}
\end{equation}where $w_{\lambda_t}^{*}$ are the optimal sparse portfolio weights for penalty parameter $\lambda_t$ and $w_t^*$ are the weights for a target portfolio -- any portfolio decision the investor desires to benchmark against.  Regret (\ref{regretdefn}) is a random variable whose uncertainty is induced from the joint posterior distribution $(\tilde{R}_t, \Theta_t \mid \textbf{R}_{t-1})$ from step 1 of the procedure.

We use the regret random variable as a tool for sparse portfolio selection.  Each portfolio decision indexed by $\lambda_t$ is assigned a number:
\begin{equation} \label{probdefn}
	\begin{split}
		\pi_{\lambda_t} = \mathbb{P}[\rho(w_{\lambda_t}^{*},w_t^*,\tilde{R}_t) < 0]
	\end{split}
\end{equation}which is the probability that the sparse portfolio is no worse than the dense (target) portfolio.  In other words, $\pi_{\lambda_t}$ is the probability that I will not ``regret" investing in the sparse $\lambda_t$-portfolio over the target portfolio.  This may also be called the \textit{satisfaction probability} for the sparse $\lambda_t$ portfolio decision.

\begin{figure}[H]
\renewcommand{\figurename}{Figure}
\centerline{
  \includegraphics[scale=.4]{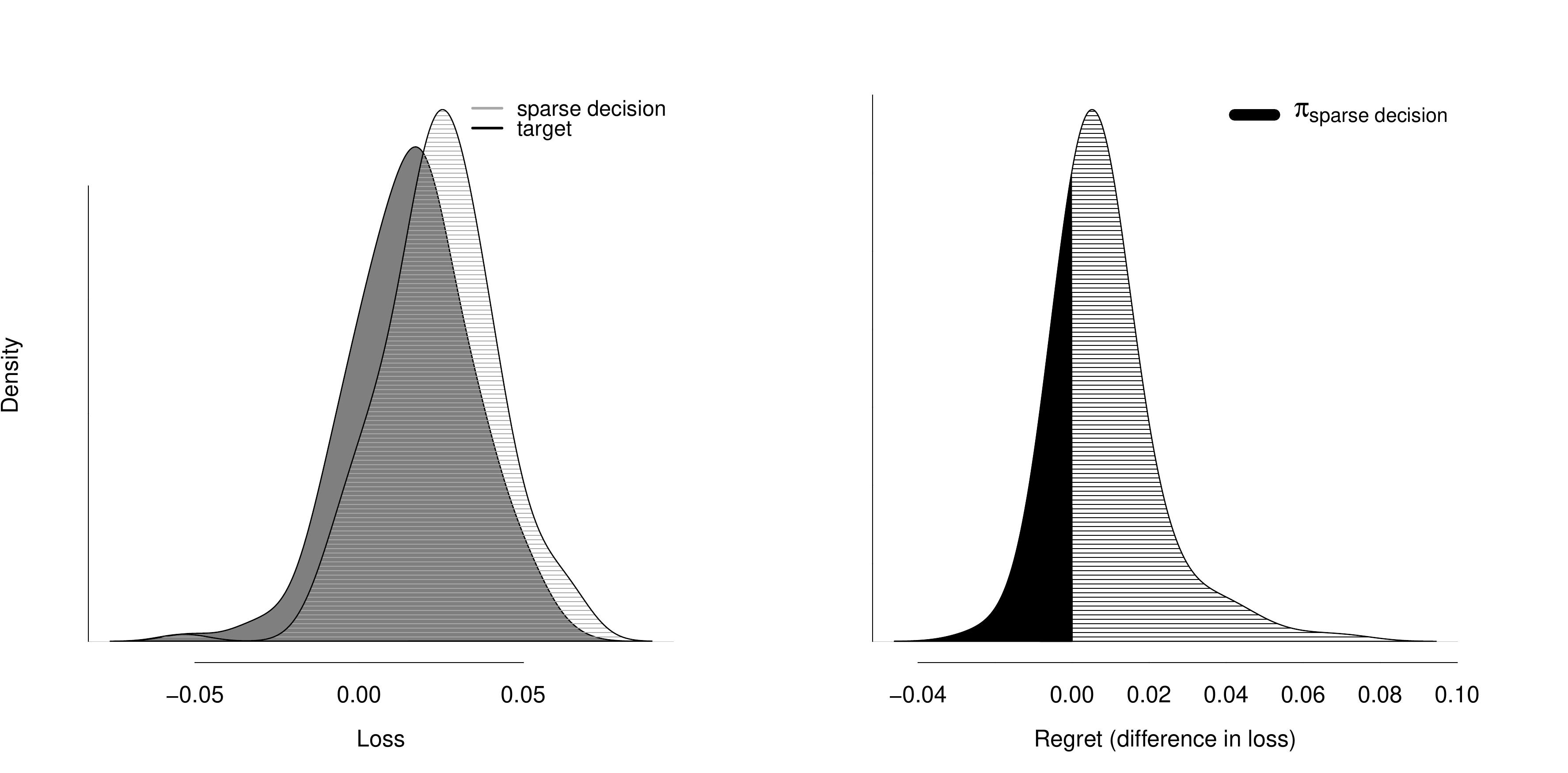}}
  \caption{Loss (left) and regret (right) for an example using returns on 25 passive indices.  The loss is defined by the log cumulative wealth. The sparse decision is a portfolio invested in 4 indices and is represented by the light shaded gray region.  The target decision is a portfolio optimized over all 25 indices and is represented by the shaded black region. The regret distributions shown on the right represent the random variables constructed by subtracting the sparse decision loss from the target loss.  Additionally, the black shaded region on the right shows $\pi_{\text{sparse decision}}$: The probability that the sparse decision is no worse than the target decision. }
  \label{deltaexamplenew}
\end{figure}

In Figure (\ref{deltaexamplenew}), we provide an illustration of the connection bewteen the loss and regret random variables.  This figure is constructed using returns on 25 passive indices and a next period ``log cumulative wealth" utility function.  This is done for a snapshot in time with a focus on one sparse decision that is invested in 4 out of the 25 indices. The investor is considering this decision versus her target -- a portfolio optimized over all 25 indices.  The left figure displays the loss distributions of the sparse decision and target. The probability mass of the sparse loss is gathered at larger values compared with the target loss.  It is ``more costly" (higher loss potential) to neglect diversification benefits and invest in fewer assets.

The right plot in Figure (\ref{deltaexamplenew}) displays the regret distribution for the sparse decision.  This is constructed by taking the difference between the sparse and target losses, as given in Equation (\ref{regretdefn}), defining the regret random variable.  With the regret distribution in hand, we can compute the probability that the sparse decision is no worse than the target portfolio given by Equation (\ref{probdefn}) -- this may be referred to as the ``satisfaction probability" for the sparse decision.  This is shown by the black shaded area on the right in Figure (\ref{deltaexamplenew}).  The larger this probability, the ``closer" the sparse decision's loss is to the target loss.  By making a decision that satisfies a lower bound on this probability called $\kappa$, we are able to control our chance of being the same or better than a target portfolio.  A lower bound ($\kappa$) on the probability of satisfaction (no regret) implies an upper bound ($1-\kappa$) on the probability of regret.

The investor's portfolio decision boils down to answering the question first posed in the introduction: \textit{With what degree of certainty do I want my simple portfolio to be no worse than the target portfolio?}  As the investor moves through time, the loss and regret distributions will evolve and so will the probability associated with the sparse $\lambda_t$ portfolio decisions.  A dynamic sparse portfolio decision extends this probability thresholding approach to a time varying framework. The investor chooses the a portfolio decision satisfying $\pi_{\lambda_t} > \kappa \hspace{3mm} \forall t$, ensuring she holds a portfolio that satisfies her regret tolerance in every investing period.
\begin{figure}[H]
\renewcommand{\figurename}{Figure}
\centering
  \includegraphics[scale=.43]{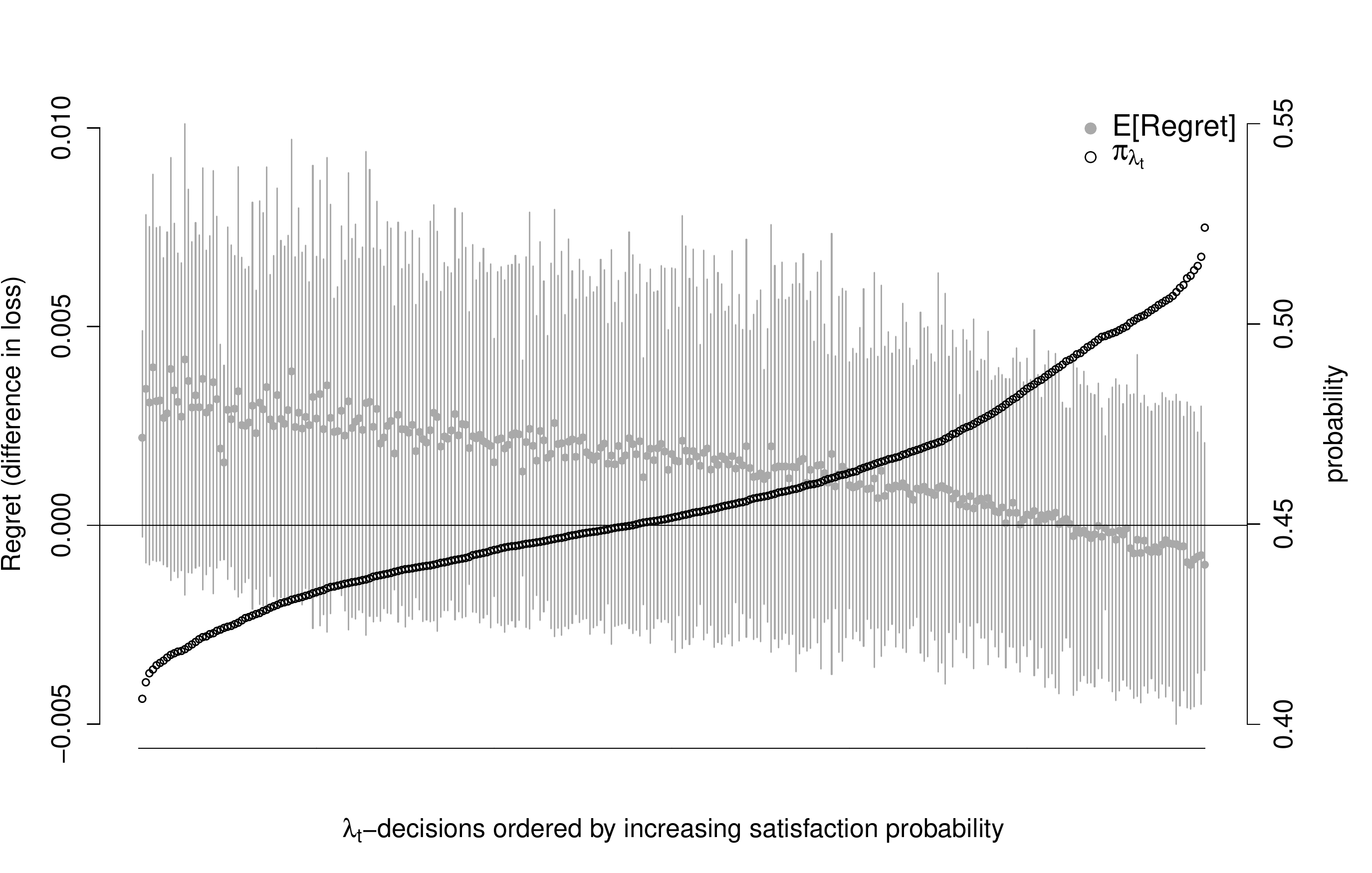}
  \caption{Regret distributions (left vertical axis) and $\pi_{\lambda_t}$ (right vertical axis) for increasing $\pi_{\lambda_t}$ values from left to right on the horizontal axis. Displayed are 300 sparse portfolio decisions indexed by $\lambda_t$. As the satisfaction probability ($\pi_{\lambda_t}$) increases, the mean regret represented by the gray dots will will typically trend downwards.  Gray bands represent 20\% centered posterior credible intervals for the regret.}
  \label{deltaexample}
\end{figure}

In Figure (\ref{deltaexample}), we show an example sequence of regret distributions (right vertical axis and gray bars) indexed by $\lambda_t$ as well as the satisfaction probability $\pi_{\lambda_t}$ (left vertical axis and open circles).  Specifically, we show the regret distributions for 300 sparse decisions under the log cumulative wealth utility and versus a target portfolio optimized over all available assets.  The highest regret decisions (by satisfaction probability) are on the left, and they become less regretful as one moves to the right on the horizontal axis. These sparse portfolio decision are all fairly close in terms of loss to the target decision.  Therefore, the corresponding regret distributions hover around zero, and the $\pi_{\lambda_t}$'s hover around 0.5.  Exploratory plots like Figure (\ref{deltaexample}) aid in choosing a proper value for the $\kappa$ threshold.  In this case $\pi_{\lambda_t} > 0.5$ is quite high.


Once the time invariant threshold $\kappa$ is specified, a dynamic selection strategy is easily implementable. At each time $t$, we are presented with a set of decisions such as those displayed in Figure (\ref{deltaexample}) and choose the sparse portfolio decision such that its $\pi_{\lambda_t} > \kappa$.  If there are several sparse decisions satisfying the threshold, we may choose the one whose $\pi_{\lambda_t}$ is closest to $\kappa$. Of course, there is flexibility in the final step of selecting among admissible sparse decisions.  For example, one may select a sparse decision at time $t$ that is ``close" (in terms of a norm or assets held) to the previous sparse decision at time $t-1$ to reduce transaction costs related to the buying and selling of assets.  These features will be discussed in the Application section.

\section{Application}

This section is divided into three parts.  First, we describe a dynamic linear model used to infer time-varying moments of asset returns and fulfill step 1 of the methodology. Second, we specify loss and resulting regret functions used for selection (steps 2 and 3).  Third, we demonstrate the methodology using a set of 25 passive funds.  In the demonstration sections, we consider a simple example and a more practical case study. 

\subsection{Model specification and data}
The general model we infer parameterizes the distribution of future asset returns with a mean and covariance indexed by time: $\tilde{R}_t \sim \Pi(\mu_t,\Sigma_t)$. An important feature of our proposed methodology is that any model providing estimations of these time varying moments may be used.  

To demonstrate our methodology, we estimate a dynamic linear model (DLM) motivated by \cite{FF5} who detail five ``risk factors" with future returns $\tilde{R}_t^{F}$ as relevant to asset pricing. Specifically, we model the joint distribution of future asset returns and factor returns $p(\tilde{R}_{t},\tilde{R}_{t}^{F})$ compositionally by modeling $p(\tilde{R}_{t} \mid \tilde{R}_{t}^{F})$ and $p(\tilde{R}_{t}^{F})$.  Following the dynamic linear model setup from \cite{harrison1999bayesian}, the future return of asset $i$ is a linear combination of future factor returns ($\tilde{R}_{t}^{i} \mid \tilde{R}_{t}^{F}$):
\begin{equation}\label{DLM1}
	\begin{split}
		\tilde{R}_{t}^{i} &= (\beta_{t}^{i})^{T}\tilde{R}_{t}^{F} + \epsilon_{t}^{i}, \hspace{6mm} \epsilon_{t}^{i} \sim N(0,1/\phi_{t}^{i}),
		\\
		\beta_{t}^{i} &= \beta_{t-1}^{i} + w_{t}^{i}, \hspace{6mm} w_{t}^{i} \sim \text{T}_{n_{t-1}^{i}}(0,W_{t}^{i}),
		\\
		\vspace{0mm}
		\\
		\beta_{0}^{i} &\mid D_{0} \sim \text{T}_{n_{0}^{i}}(m_{0}^{i},C_{0}^{i}),
		\\
		\phi_{0}^{i} &\mid D_{0} \sim \text{Ga}(n_{0}^{i}/2,d_{0}^{i}/2),
		\\
		\vspace{0mm}
		\\
		\beta_{t}^{i} &\mid D_{t-1} \sim \text{T}_{n_{t-1}^{i}}(m_{t-1}^{i},R_{t}^{i}), \hspace{6mm} R_{t}^{i} = C_{t-1}^{i}/\delta_{\beta},
		\\
		\phi_{t}^{i} &\mid D_{t-1} \sim \text{Ga}(\delta_{\epsilon} n_{t-1}^{i}/2,\delta_{\epsilon} d_{t-1}^{i}/2),
	\end{split}
\end{equation}where $W_{t}^{i} = \frac{1-\delta_{\beta}}{\delta_{\beta}}C_{t-1}^{i}$ and $D_{t}$ is all information up to time $t$.  This model permits the coefficients on the factors as well as the observation and state level variances to vary in time.  Pre-specified discount factors $\delta_{\epsilon}$ and $\delta_{\beta}$ $\in (0.8,1)$ accomplish this goal for the observation and state level variances, respectively.  Also, note that $C_{t}^{i}$ (the posterior variance of the state equation for $\beta_{t}^i$) is updated through moments of the prior $\beta_{t}^{i} \mid D_{t-1}$ and the one-step ahead forecast distribution $\tilde{R}_{t}^{i} \mid D_{t-1}$. Theorem 4.1 in \cite{harrison1999bayesian} provides the general updating equations for the univariate DLM.  Table 10.4 in the book summarizes the recurrence relationships in the special case of variance discounting, providing the moments of the posteriors of the parameters $\{m_t^i,C_t^i,n_t^i,d_t^i\}$ for all $t$ and each asset $i$.

We model the five factor future returns $\tilde{R}_{t}^{F}$ with a full residual covariance matrix using the following matrix normal dynamic linear model:
\begin{equation}\label{DLM2}
	\begin{split}
		&\tilde{R}_{t}^{F} = \mu_{t}^{F} + \nu_{t} \hspace{10mm} \nu_{t} \sim \text{N}(0,\Sigma_{t}^{F}), 
		\\
		&\mu_{t}^{F} = \mu_{t-1}^{F} + \Omega_{t} \hspace{6mm} \Omega_{t} \sim \text{N}(0,W_{t},\Sigma_{t}^{F}),
		\\
		\vspace{0mm}
		\\
		&(\mu_{0}^{F},\Sigma_{0}^{F} \mid D_{0}) \sim \text{NW}_{n_{0}}^{-1}(m_{0},C_{0},S_{0}),
		\\
		\vspace{0mm}
		\\
		&( \mu_{t}^{F}, \Sigma_{t}^{F} \mid D_{t-1}) \sim \text{NW}_{\delta_{F}n_{t-1}}^{-1}(m_{t-1},R_{t},S_{t-1}),
		\\
		&R_{t} = C_{t-1}/\delta_{c},
	\end{split}
\end{equation}where $W_{t} = \frac{1-\delta_{c}}{\delta_{c}}C_{t-1}$.  Analogous to Model (\ref{DLM1}), the discount factors $\delta_{F}$ and $\delta_{c}$ in Model (\ref{DLM2}) serve the same purpose of permitting time variation in the observation and state level variances, respectively. An added benefit of (\ref{DLM2}) is that $\Sigma_{t}^{F}$ is a full residual covariance matrix.  

Elaborating on the intuition behind Models (\ref{DLM1}) and (\ref{DLM2}) and guided by \cite{harrison1999bayesian}, the purpose of variance discounting is to provide a natural way to evolve the variance from the posterior to the prior while maintaining conjugacy for sequential updating.  For example, consider the posterior of the precision in Model (\ref{DLM1}):
\begin{equation} \label{phiexample}
	\begin{split}
		\phi_{t-1}^i \mid D_{t-1} \sim \text{Ga}(n_{t-1}^{i}/2,d_{t-1}^{i}/2).
	\end{split}
\end{equation}To construct $p(\phi_{t}^i \mid D_{t-1})$, we wish to maintain a Gamma form so it is conjugate with the likelihood function for $\tilde{R}_t^i$ given by the one-step ahead forecast distribution. One reasonable approach is to preserve the mean of Distribution (\ref{phiexample}), but inflate the variance by discounting the degrees of freedom parameter $n_{t-1} \rightarrow \delta_{\epsilon}n_{t-1}$.  The prior distribution then becomes:
\begin{equation}\label{phiexample2}
	\begin{split}
		\phi_{t}^i \mid D_{t-1} \sim \text{Ga}(\delta_{\epsilon}n_{t-1}^{i}/2,\delta_{\epsilon}d_{t-1}^{i}/2).
	\end{split}
\end{equation}Moving from Distribution (\ref{phiexample}) to (\ref{phiexample2}) increases the dispersion of the prior to represent a ``loss of information" characteristic of moving forward to time $t$ with a lack of complete information in $D_{t-1}$.  The remaining discount factors $\delta_{\beta},\delta_{C}$, and $\delta_{F}$ in Models (\ref{DLM1}) and (\ref{DLM2}) serve the analogous purpose of \textit{variance inflation} for their respective stochastic processes.

The limiting behavior of variance-discounted learning corresponds to exponentially weighting historical data with decreasingly smaller weights given to values further in the past (see sections 10.8.2-3 in \cite{harrison1999bayesian}).  Larger discount factors correspond to slower decaying weights and suggest the time series of parameters is slower-fluctuating.  Smaller discount factors intrinsically mean we have less data with which to estimate the parameters because the sequence is believed to be  more rapidly fluctuating.  Thus, the choice of discount factors amounts to choosing decaying weights for previous data that are relevant for predicting the parameters today.

Models (\ref{DLM1}) and (\ref{DLM2}) together constitute a time-varying model for the joint distribution of future asset and factor returns: $p(\tilde{R}_{t},\tilde{R}_{t}^{F}) = p(\tilde{R}_{t} \mid \tilde{R}_{t}^{F})p(\tilde{R}_{t}^{F})$. As detailed in \cite{harrison1999bayesian}, they are Bayesian models that have closed form posterior distributions of all parameters at each time $t$, and the absence of MCMC is convenient for fast updating and uncertainty characterization -- a necessary ingredient for our regret-based portfolio selection procedure.  Under these models, we obtain the following mean and covariance structure:
\begin{equation}\label{moments}
	\begin{split}
		\mu_{t} &= \beta_{t}^{T}\mu_{t}^{F},
		\\
		\Sigma_{t} &= \beta_{t} \Sigma_{t}^{F} \beta_{t}^{T} + \Psi_{t},
	\end{split}
\end{equation}where column $i$ of $\beta_{t}$ are the coefficients on the factors for asset $i$, $\beta_{t}^{i}$.  Also, $\Psi_{t}$ is a diagonal matrix with $i$th element $\Psi_{tii} = 1/\phi_{t}^{i}$.  The parameters $\Theta_t = (\mu_t,\Sigma_t)$ are inputs to step 2 of the procedure.
%

\subsubsection{Data and choice of discount factors}

We use data on the 25 most highly traded (i.e., most liquid) equity funds from ETFdb.com as our investable assets. This is monthly data from the Center for Research in Security Prices (CRSP) database from February 1992 through May 2016 \citep{CRSP}.  If the fund's inception date is after 1992, we backfill its return data with the index it is mandated to track.

The fund names, tickers, and sample statistics are displayed in Table (\ref{tablepassivelist}).  The returns on the Fama-French five factors are obtained from the publicly available data library on Ken French's website.\footnote{http://mba.tuck.dartmouth.edu/pages/faculty/ken.french/}  We start with 10 years of data to train the model and begin forming portfolios in February 2002.  
\begin{table}[H]
{\footnotesize  {\centering
\begin{tabular}{lccc}
\textbf{Fund name} & \textbf{Ticker} & \textbf{Return (\%)} & \textbf{St. Dev. (\%)} \\
  \midrule
SPDR Dow Jones Industrial Average & DIA & 8.06 & 14.15 \\ 
SPDR S\&P 500 ETF & SPY & 7.46 & 14.34 \\
SPDR Industrial Select Sector & XLI & 9.19 & 16.49 \\
Guggenheim S\&P 500 Equal Weight ETF & RSP & 9.37 & 15.92 \\
iShares MSCI Emerging Markets & EEM & 6.06 & 22.78 \\ 
iShares MSCI EAFE & EFA & 3.94 & 16.43 \\  
iShares MSCI Germany & EWG & 6.82 & 22.23 \\  
iShares MSCI Japan & EWJ & 0.22 & 19.28 \\ 
iShares MSCI United Kingdom & EWU & 4.63 & 15.88 \\ 
iShares MSCI South Korea Capped & EWY & 9.46 & 36.78 \\
iShares MSCI Eurozone & EZU & 6.00 & 19.85 \\
iShares S\&P 500 Value & IVE & 7.36 & 14.95 \\
iShares Core S\&P 500 & IVV & 7.48 & 14.34 \\
iShares Russell 1000 & IWB & 8.22 & 14.00 \\
iShares Russell 1000 Value & IWD & 8.10 & 14.26 \\
iShares Russell 1000 Growth  & IWF & 7.63 & 14.56 \\
iShares Russell 2000 & IWM & 8.00 & 18.76 \\
iShares Russell 2000 Value & IWN & 7.78 & 18.67 \\ 
iShares Russell 2000 Growth & IWO & 6.63 & 22.14 \\
iShares Russell Mid-Cap Growth & IWP & 8.52 & 19.98 \\
iShares Russell Mid-Cap & IWR & 9.52 & 15.96 \\
iShares Russell 3000 & IWV & 7.52 & 14.61 \\
iShares US Real Estate & IYR & 9.32 & 19.12 \\ 
iShares US Technology & IYW & 11.64 & 26.09 \\
iShares S\&P 100 & OEF & 7.32 & 14.61 \\
  \midrule
\end{tabular}
}}
  \vspace{3mm}
   \caption{List of exchange-traded funds (ETFs) used for the empirical study. Also displayed are the ticker symbols and realized return and standard deviation (annualized) over their sample period.}\label{tablepassivelist}
\end{table}

Step 1 in our procedure is specifying and inferring a model for asset returns.  For our empirical analysis, we use Models (\ref{DLM1}) and (\ref{DLM2}). The discount factors for the factor coefficient and factor mean processes are set to $\delta_{c} = \delta_{\beta} = 0.9925$, and we consider time varying residual variances $\delta_{F} =\delta_{\epsilon}=0.97$.  Evidence of time varying residual variance is well-documented in the finance literature (see for example, \cite{ng1991tests}).  The discount factors are chosen to incorporate an adequate amount of data in the exponentially weighted moving window.  When $\delta = 0.9925$ ($0.97$), data eight (three) years in the past receives half the weight of data today.  We require slower decay for the factor coefficients and mean processes because more data is needed to learn these parameters than residual volatility.

\subsection{Loss and regret specification}

%

We consider a loss function defined by the negative log cumulative return of a portfolio decision for $N$ assets.  Recalling general form of the loss function: $\textbf{L}_{\lambda_t}(w_t,\tilde{R}_t) = \mathcal{L}(w_t,\tilde{R}_t) + \Phi(\lambda_t,w_t)$, define:
\begin{equation} \label{lossorig}
	\begin{split}
		\mathcal{L}(w_t,\tilde{R}_t) &= - \log \left( 1 + \sum_{k=1}^{N} w_t^k\tilde{R}_t^k \right),
	\end{split}
\end{equation}The utility in Loss (\ref{lossorig}) may be viewed as a version of the Kelly portfolio criterion \citep{kelly1956new} where the investor's preferences involve the portfolio growth rate.  The complexity function $\Phi(\lambda_t,w_t)$ is separately specified in each of the two examples to follow.


Portfolio decisions $w_t$ may now be evaluated using the negative log cumulative return preferences given by $\mathcal{L}(w_t,\tilde{R}_t)$.  However, in order to find these portfolio decisions, we must first optimize the expectation of Loss (\ref{lossorig}).  We do this in two steps: First, we approximate the loss using a second order Taylor expansion, and second, we take the expectation over all unknowns and optimize for each $\lambda_t$.  

Following the work of \cite{rising2012partial}, we consider a convenient second order Taylor approximation $\mathcal{L}(w_t,\tilde{R}_t) \approx \mathring{\mathcal{L}}(w_t,\tilde{R}_t) $ of the original Loss (\ref{lossorig}) expanded about $w_{0} = \vec{0}$:
\begin{equation}
	\begin{split}
		\mathring{\mathcal{L}}(w_t,\tilde{R}_t) &=  \frac{1}{2}\sum_{k=1}^{N}\sum_{j=1}^{N} w_t^kw_t^j\tilde{R}_t^k\tilde{R}_t^j - \sum_{k=1}^{N} w_t^k\tilde{R}_t^k,
	\end{split}
\end{equation}where we write the approximate loss including the penalty function as $\mathring{\textbf{L}}_{\lambda_t}(w_t,\tilde{R}_t) = \mathring{\mathcal{L}}(w_t,\tilde{R}_t) + \Phi(\lambda_t,w_t)$.  The approximate expected loss $\mathbb{E}[\mathring{\textbf{L}}_{\lambda_t}(w_t,\tilde{R}_t)]$ is written as:
\begin{equation}
	\begin{split}
		\mathbb{E}[\mathring{\textbf{L}}_{\lambda_t}(w_t,\tilde{R}_t)] = \mathbb{E}\left[  \frac{1}{2}\sum_{k=1}^{N}\sum_{j=1}^{N} w_t^kw_t^j\tilde{R}_t^k\tilde{R}_t^j - \sum_{k=1}^{N} w_t^k\tilde{R}_t^k + \Phi(\lambda_t,w_t)  \right].
	\end{split}
\end{equation}

With the posterior distribution $(\Theta_t, \tilde{R}_t \mid \textbf{R}_{t-1})$ in hand from step 1, we can take the expectation. We integrate over $(\tilde{R}_t \mid \Theta_t, \textbf{R}_{t-1})$ followed by $(\Theta_t \mid \textbf{R}_{t-1})$ to obtain the integrated approximate loss function: 
\begin{equation} \label{integratedloss}
	\begin{split}
		\mathring{\textbf{L}}_{\lambda_t}(w_t) = \mathbb{E}[\mathring{\textbf{L}}_{\lambda_t}(w_t,\tilde{R}_t)] & = \mathbb{E}_{\Theta_t}\left[\mathbb{E}_{\tilde{R}_t \mid \Theta_t}\left[\mathring{\textbf{L}}_{\lambda_t}(w_t,\tilde{R}_t)\right]\right]
		\\
		  & = \mathbb{E}_{\Theta_t}\left[\frac{1}{2}w_t^{T}{\Sigma}_t^{\text{\tiny nc}} w_t - w_t^{T}\mu_t + \Phi(\lambda_t,w_t)\right]
		\\
		& = \frac{1}{2}w_t^{T}\overline{\Sigma}_t^{\text{\tiny nc}} w_t - w_t^{T}\overline{\mu}_t + \Phi(\lambda_t,w_t),
	\end{split}
\end{equation}where the overlines denote posterior means of the mean $\mu_t$ and  non-central second moment $\Sigma_t^{\text{\tiny nc}}$.  The non-central second moment is calculated from the variance as $\Sigma_t^{\text{\tiny nc}} = \Sigma_t + \mu_t\mu_t^T$.  Loss function (\ref{integratedloss}) may be minimized for a range of $\lambda_t$ values at each time $t$.

In the subsections to follow, we will present two analyses using this model and data.  First, we discuss a simple unsupervised example to demonstrate the regret-based selection procedure.  Second, we present an in depth practical case study.

\subsection{Simple example: Portfolio decisions with limited gross exposure}

In this section, we present a simple example demonstrating the main components of regret-based selection.  We complete Loss function (\ref{overallloss}) by specifying the complexity function as the $\ell_1$ norm of the weight vector: $\Phi(\lambda_t,w_t) = \lambda_t \norm{ w_t }_1$. The complexity function measures \textit{gross exposure} of a decision by summing the absolute value of each position: $\norm{ w_t }_{1} = \Sigma_i | w_t^i |$.  Decisions with larger absolute value components will evaluate to larger $\ell_1$ norms. The penalty parameter $\lambda_t$ corresponds directly to a single portfolio decision by amplifying the penalty in the loss function.

The approximate Loss function (\ref{integratedloss}) is now convex and may be written as:
\begin{equation} \label{integratedlossconvex}
	\begin{split}
		\mathring{\textbf{L}}_{\lambda_t}(w_t) = \frac{1}{2}w_t^{T}\overline{\Sigma}_t^{\text{\tiny nc}} w_t - w_t^{T}\overline{\mu}_t + \lambda_t \norm{ w_t }_1.
	\end{split}
\end{equation}Loss function (\ref{integratedlossconvex}) is readily optimized by a variety of software packages -- please see Appendix (\ref{applossfunopt}) for details.  Given its computational convenience, it possesses a couple important features worth noting.

First, Loss (\ref{integratedlossconvex}) requires no enumeration of decisions; it can be minimized quickly for a range of $\lambda_t$.  In this way, it is an ``unsupervised" approach to the sparse portfolio selection problem. Second, $\lambda_t$ now has explicit meaning beyond indexing the decisions. Since it is multiplying the complexity function, larger (smaller) $\lambda_t$ will generally correspond to sparser (denser) portfolio decisions.  Conveniently, this displays the regret-based procedure's usefulness in selecting tuning parameters in penalized optimization problems with time-varying inputs.  The dynamic nature of asset return data renders traditional cross validation approaches using i.i.d. sampled testing and training splits inappropriate.

\begin{figure}[H]
\renewcommand{\figurename}{Figure}
\centering
  \includegraphics[scale=.43]{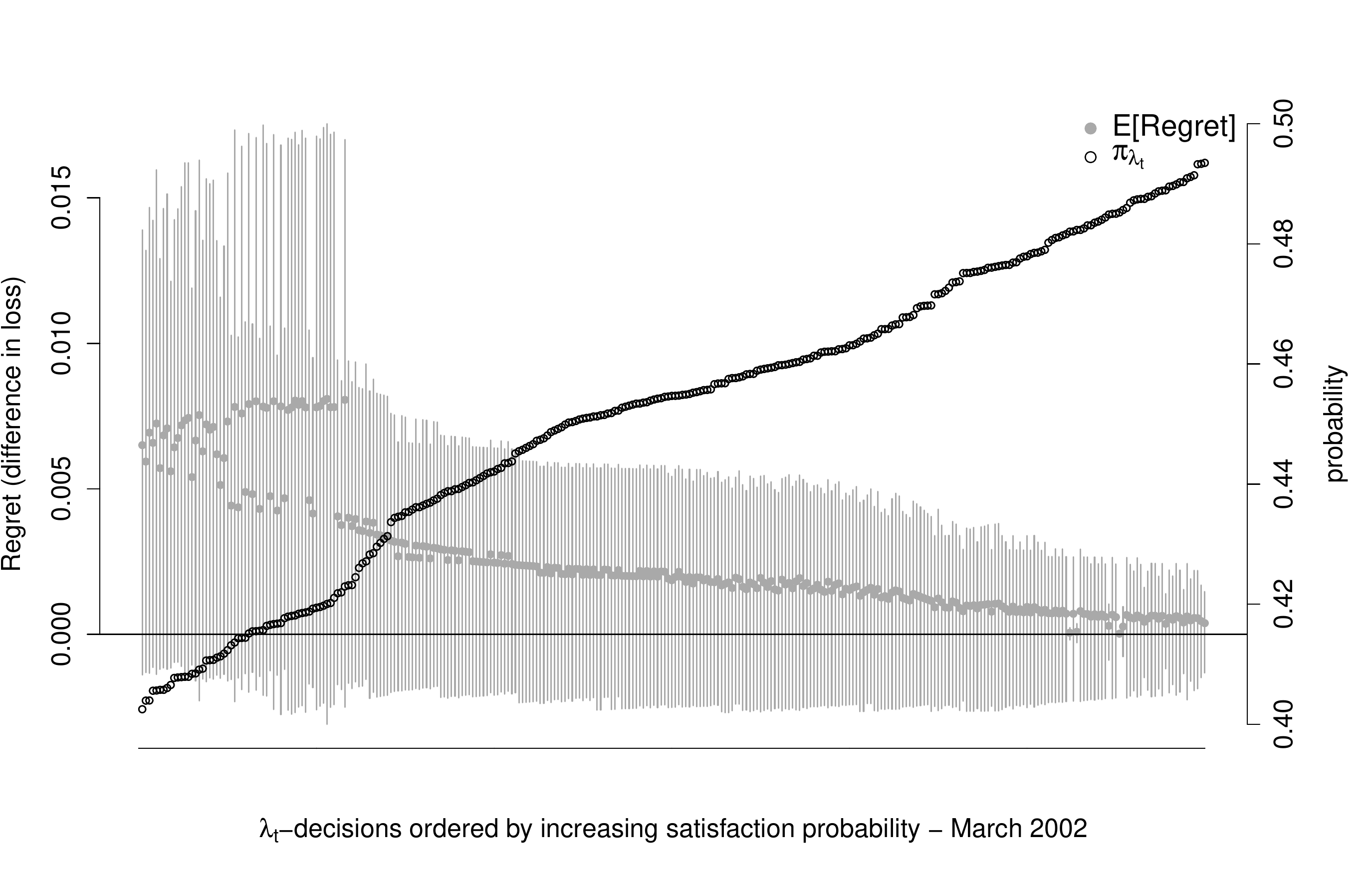}
  \caption{Regret distributions (left vertical axis) and $\pi_{\lambda_t}$ (right vertical axis) for increasing $\pi_{\lambda_t}$ values from left to right on the horizontal axis. Displayed are 300 of the sparse portfolio decisions indexed by $\lambda_t$ for March 2002. As the satisfaction probability ($\pi_{\lambda_t}$) increases, the mean regret represented by the gray dots will will typically trend downwards.  Gray bands represent 20\% centered posterior credible intervals for the regret.}
  \label{deltaexampleLASSO}
\end{figure}

%

We optimize Loss function (\ref{integratedlossconvex}) for a range of $\lambda_t$ and for each time $t$.  Specifically, we consider 500 $\lambda_t$ values spanning the sparsest ``one asset decision" to the $\lambda_t=0$ decision. The latter decision is referred to as the unpenalized \textit{Kelly optimal} portfolio and puts nonzero weight in all available assets. This is used as the target decision since there is no limit on gross exposure and will also be called the \textit{dense portfolio}.  We also normalize all decisions to sum to one, and allow long \textit{and} short positions in assets.  We normalize all decisions so that the investor is neither borrowing nor investing in the risk free (cash) rate. Instead, she is fully invested in a risky asset portfolio.  In other words, denoting $w_{\text{cash}}$ and $w_{\text{risky}}$ as the percentage of wealth in cash and risky assets (the ETFs) respectively, $w_{\text{cash}}+w_{\text{risky}}=1$ will always hold, and this paper considers the case when $w_{\text{risky}}=1$.

At each point in time, the investor would like to choose among the 500 sparse decisions indexed by $\lambda_t$.  This may be done by first computing the corresponding satisfaction probabilities $\pi_{\lambda_t}$ for each of the 500 decisions under consideration and then choosing one that satisfies and a pre-specified threshold $\kappa$.  Recall that $\pi_{\lambda_t}$ is defined in (\ref{probdefn}) as the probability that the regret (versus the dense target) is less than zero.  The utility, as specified in Equation (\ref{lossorig}), is the next period log cumulative wealth.

Figure (\ref{deltaexampleLASSO}) displays the cross-sectional regret distributions (left vertical axis) and satisfaction probabilities $\pi_{\lambda_t}$ (right vertical axis) for 300 of the sparse decisions in March 2002.  As prescribed by the regret-based procedure, the investor uses this information to select a portfolio. The satisfaction probabilities span 0.4 to 0.5 indicating that the decisions in this investing period are all quite similar. Guided by this figure, we choose a $\kappa=42.5\%$ threshold to construct an example sequence of selected portfolios. 

Once the static threshold is selected, we can iterate the selection procedure through time.  At each time $t$, the investor is confronted with ex ante regret information provided by a cross-sectional plot like Figure (\ref{deltaexampleLASSO}) and selects a portfolio that satisfies the threshold $\kappa$.  Once the sequence of decisions is constructed, we can look at how the regret distribution varies \textit{over time}.

\begin{figure}[H]
\renewcommand{\figurename}{Figure}
\centerline{
  \includegraphics[scale=.43]{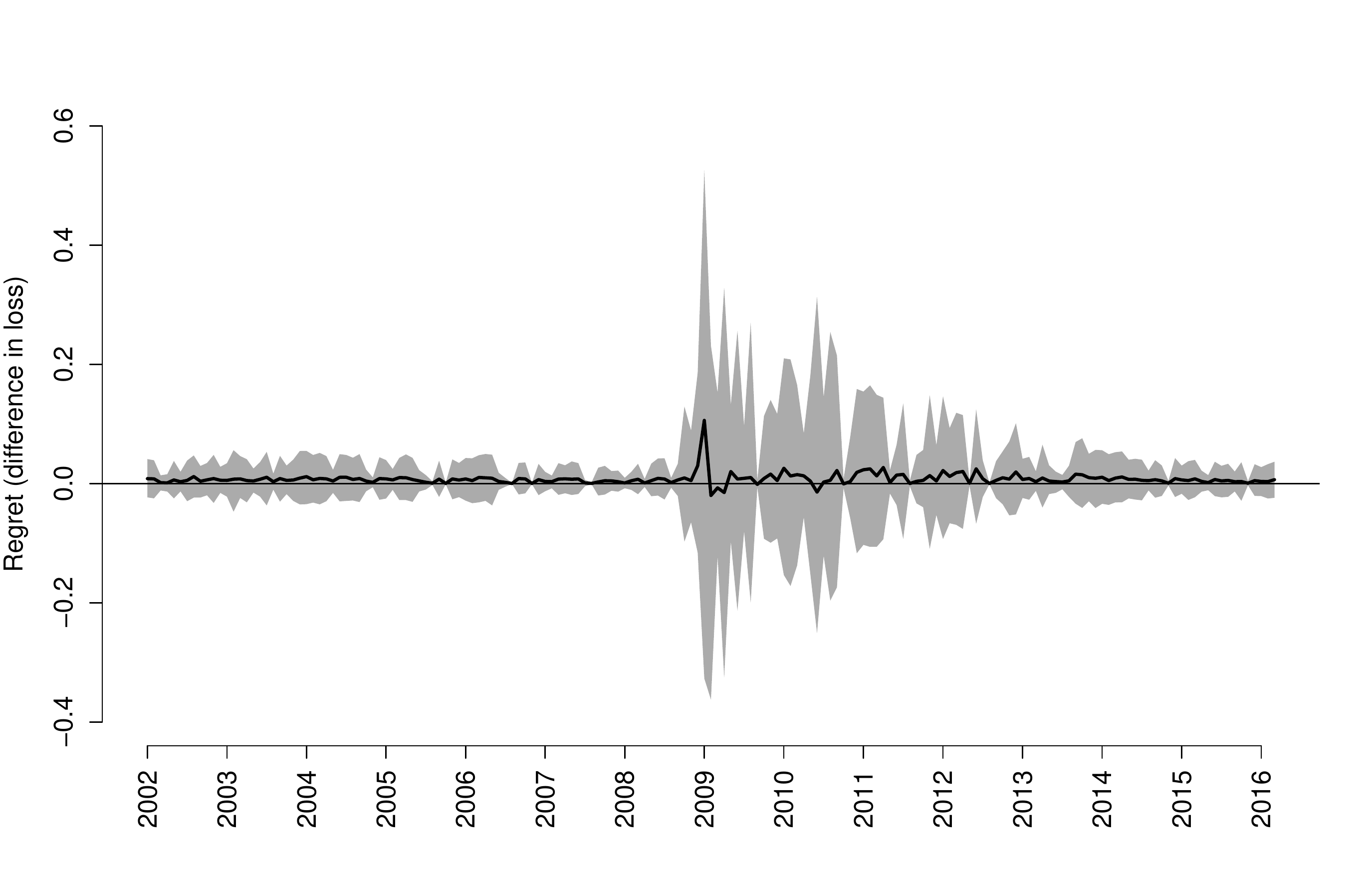}}
  \caption{The evolution of the ex ante regret distributions for the sparse long/short portfolio decision given by a $\kappa=42.5\%$ threshold and versus the unpenalized Kelly optimal target.  The mean regret is shown by the lines and the surrounding areas represent the evolving centered 60\% posterior credible intervals.}
  \label{regretevolveLASSO}
\end{figure}

Figure (\ref{regretevolveLASSO}) shows precisely how the regret of the selected decisions evolve over time.  This example demonstrates how both the mean (black line) and variance (surrounding shaded black regions) of regret can vary substantially. Notice that the regret is close to zero with small variance for most periods of time.  However, surrounding the financial crisis in 2009, the mean increases and then drops below zero and the variance increases.  When regret is negative, the utility of sparse portfolio decision exceeds that of the dense portfolio.  During crisis periods shortly into 2009, sparse portfolio decisions appear to be preferred (as measured by ex ante investor utility) to the dense portfolio.  Nonetheless, this drop in mean is accompanied by increased variance which informs the investor to be wary of the precision of her regret assessment.

\subsection{Case study: Selection among a large set of enumerated decisions}


The purpose of the following case study is to demonstrate regret-based portfolio selection for a layman investor.  We assume our investor would like to hold a broad market fund (SPY) and a couple more diversifying positions in other funds.  Additionally, we consider a scenario where the investor cannot hold negative positions in funds; i.e., short selling is prohibited.  Therefore, we consider decisions of only positive weights: $w_t \geq 0$ $\forall t$.  We construct a set of portfolio decisions for a layman investor using the following rules for a sparse portfolio with $q < N$ funds:

\begin{enumerate}
	\item $\geq 25\%$ of the portfolio is invested in SPY, a broad market fund tracking an index comprised of the 500 largest US companies by market capitalization.
	\item $\geq 25\%$ of the portfolio is diversified across the $q-1$ non-market funds in the following way: The $q-1$ non-market funds each have weights $\geq \frac{25}{q-1}\%$.
\end{enumerate}We consider portfolios of two, three, four, and five funds, all of which include SPY.  Each of these sparse portfolios are optimized using the unpenalized \textit{Kelly optimal} loss as the objective (Loss (\ref{integratedloss}) without the complexity function) and constraints defined as above.  Since our data has 24 funds excluding SPY, enumeration of decisions in this way results in $\Sigma_{i=1}^4 {24 \choose i} = 12,950$ sparse portfolios to select among.  Enumeration of sparse decisions implies a complexity function that measures the \textit{cardinality} or number of funds included in a portfolio.  Since the complexity function is now implicit in the sparse enumeration, $\lambda_t$ may be thought of as a convenient indexing of each possible portfolio decision.


As presented in the initial example, we must specify a target decision which then defines the regret random variable defined in Equation (\ref{regretdefn}).  We consider two targets at opposite ends of the sparsity spectrum for the empirical analysis.  

\begin{enumerate}
	\item \textbf{Dense target}: The unpenalized \textit{Kelly optimal} decision; a portfolio optimized over all available assets. Define the Kelly optimal decision as $w_t^* = \arg\min_{w_t \geq 0} \mathring{\mathcal{L}}(w_t)$ where $\mathring{\mathcal{L}}(w_t) = \mathbb{E}[\mathring{\mathcal{L}}(w_t,\tilde{R}_t)] = \frac{1}{2}w_t^{T}\overline{\Sigma}_t^{\text{\tiny nc}} w_t - w_t^{T}\overline{\mu}_t$; the optimal decision in absence of the penalty function. This is the same target used in the ``$\ell_1$ penalty" example presented above, now with a positivity constraint on the weights. 
	\item \textbf{Sparse target}: The \textit{Market}; a portfolio composed of one asset representing broad exposure to the financial market.  We choose SPY as the market fund.
\end{enumerate}

The choice of each target will give an investor vastly different perspectives on sparse portfolio selection. In the dense target case, the investor desires a sparse portfolio decision that is close (in terms of regret) to a potentially unattainable decision involving all possible funds. The sparse target turns this approach on its head.  In this case, the sparse target approach will inform the investor of the added benefit (if any) in diversifying away from a broad market fund.

\begin{figure}[H]
\renewcommand{\figurename}{Figure}
\centering
  \includegraphics[scale=.43]{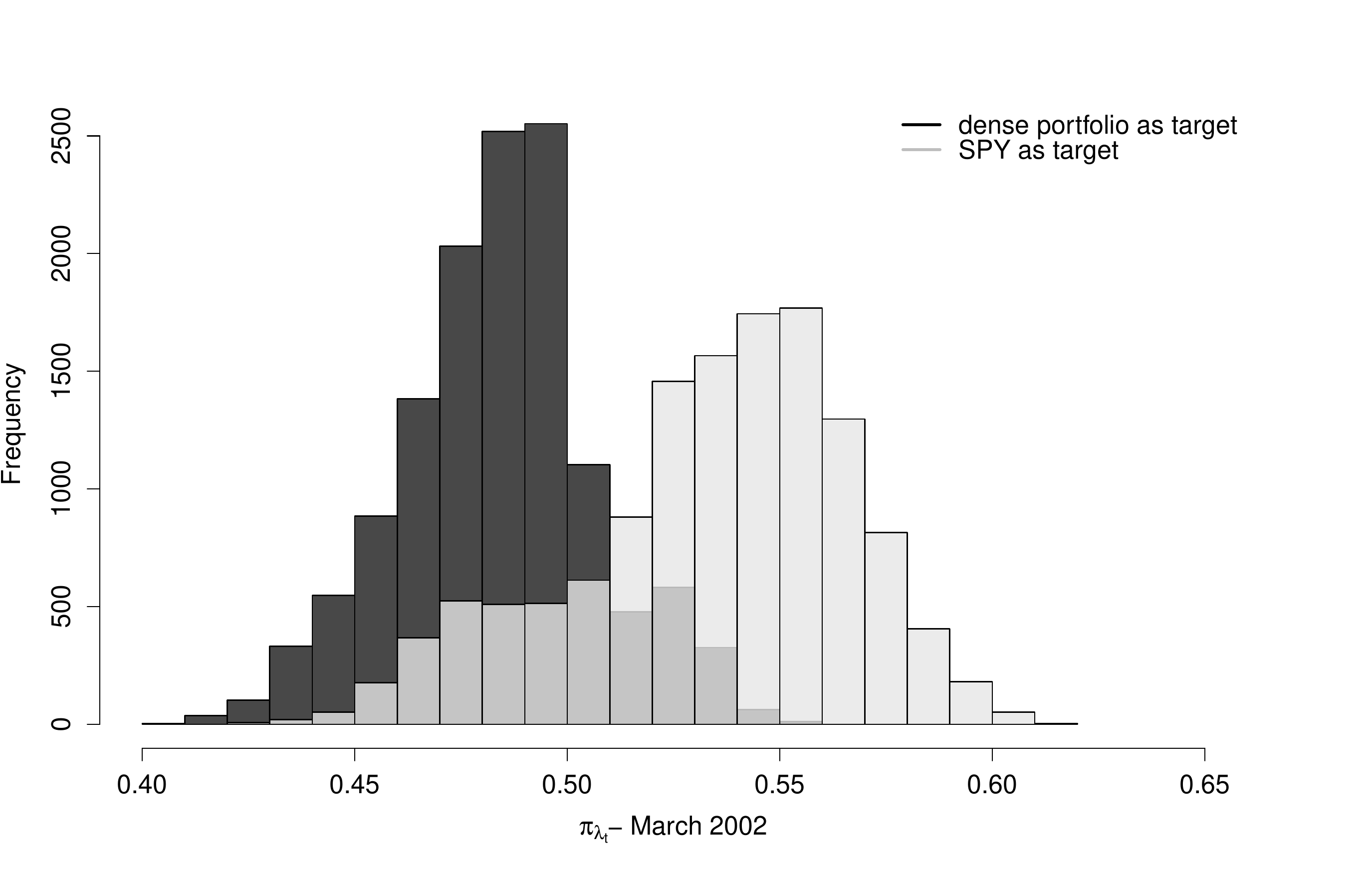}
  \caption{Histograms of the satisfaction probabilities ($\pi_{\lambda_t}$) for two target decisions: The dense portfolio (black) and SPY (i.e., sparse target, in gray).  These are shown for the March 2002 investing period and are distributions across all 12,950 enumerated sparse decisions.}
  \label{pilamexample}
\end{figure}

Each of the 12,950 sparse decisions has a probability of satisfaction versus a target ($\pi_{\lambda_t}$) which can be readily calculated via simulation at each point in time using Equations (\ref{regretdefn}) and (\ref{probdefn}) and the distribution of future returns given by Models (\ref{DLM1}) and (\ref{DLM2}). In Figure (\ref{pilamexample}), we show histograms of the satisfaction probabilities for March 2002 across all 12,950 sparse decisions.  It is related to Figure (\ref{deltaexample}) in that the satisfaction probabilities corresponding to the right vertical axis are shown in histogram form, now for various targets.  The probabilities versus the dense (SPY) target are shown in black (gray).  The dense target is the \textit{dense} portfolio optimized over all 25 funds. Satisfaction versus this dense portfolio decision are gathered at smaller probabilities when compared with the SPY portfolio decision.  Of course, the satisfaction rate versus a diversified dense portfolio will intuitively be lower than versus a sparse portfolio of a single fund.

Figure (\ref{pilamexample}) aids in the proper choice of the regret threshold $\kappa$.  When evaluated under the next period cumulative wealth utility, all long only portfolio decisions are somewhat similar.  Thus, the regret (difference in loss) will generally be gathered around values close to zero, and the satisfaction probabilities will be gathered around value close to 0.5.  As a next step, we select a $\kappa$ and present the resulting dynamic portfolio decision for the dense target and SPY target. 

\begin{table}[H]

\footnotesize
\centerline{
\centering
\begin{tabular}{K{0.8cm}|K{0.2cm}Q{0.2cm}K{0.2cm}Q{0.2cm}K{0.2cm}Q{0.2cm}K{0.2cm}Q{0.2cm}K{0.2cm}Q{0.2cm}K{0.2cm}Q{0.2cm}K{0.2cm}Q{0.2cm}K{0.2cm}Q{0.2cm}K{0.2cm}Q{0.2cm}K{0.2cm}Q{0.2cm}K{0.2cm}Q{0.2cm}K{0.2cm}Q{0.2cm}K{0.2cm}Q{0.2cm}K{0.2cm}Q{0.2cm}K{0.2cm}Q{0.2cm}K{0.2cm}Q{0.2cm}K{0.2cm}Q{0.2cm}K{0.2cm}Q{0.2cm}K{0.2cm}Q{0.2cm}K{0.2cm}Q{0.2cm}}
\textbf{Dates} & \multicolumn{2}{c}{SPY} &  \multicolumn{2}{c}{EZU} &  \multicolumn{2}{c}{EWU} &  \multicolumn{2}{c}{EWY} &  \multicolumn{2}{c}{EWG} &  \multicolumn{2}{c}{EWJ} &  \multicolumn{2}{c}{OEF} &  \multicolumn{2}{c}{IVV} &  \multicolumn{2}{c}{IVE} &  \multicolumn{2}{c}{EFA} &  \multicolumn{2}{c}{IWP} \\
  \midrule
2003 & 25 & 75 & - & - & 58 & - & - & - & - & - & - & - & - & 8.3 & - & - & - & - & - & - & - & 8.3 \\ 
  2004 & 25 & 75 & - & - & 43 & - & - & - & - & - & 20 & - & - & - & 6.2 & - & - & - & - & - & - & - \\ 
  2005 & 25 & 75 & - & - & 25 & - & - & - & 6.2 & - & 13 & - & - & 8.3 & - & - & - & - & - & - & - & - \\ 
  2006 & 62 & 75 & - & - & - & - & - & - & 6.2 & - & 19 & - & - & 12 & - & - & - & - & - & - & - & - \\ 
  2007 & 75 & 75 & - & - & - & - & 25 & - & - & 8.3 & - & - & - & - & - & - & - & - & - & - & - & 8.3 \\ 
  2008 & 44 & 75 & - & - & - & - & - & 12 & 8.3 & 13 & 21 & - & - & - & - & - & - & - & 26 & - & - & - \\ 
  2009 & 30 & 45 & - & - & - & - & 6.2 & - & - & - & 41 & - & - & - & - & 34 & - & - & 17 & 21 & 6.3 & - \\ 
  2010 & 75 & 55 & - & - & - & - & 8.3 & - & - & - & - & - & - & - & - & 26 & - & - & - & 11 & 8.3 & - \\ 
  2011 & 58 & 57 & - & - & 25 & - & - & - & - & - & - & - & - & - & - & 26 & - & - & - & - & 8.3 & - \\ 
  2012 & 29 & 25 & 8.3 & - & - & - & - & - & - & - & 54 & - & - & 56 & - & 6.2 & - & - & - & - & - & - \\ 
  2013 & 34 & 25 & - & - & - & - & - & 6.2 & - & 6.2 & 49 & - & - & 56 & - & - & - & - & - & - & 8.3 & - \\ 
  2014 & 25 & 75 & - & - & - & - & - & - & - & - & 37 & - & 26 & - & - & - & - & - & 6.2 & - & - & - \\ 
  2015 & 45 & 25 & - & - & - & - & - & - & - & - & 39 & 36 & - & 27 & - & - & 8.3 & - & - & 6.2 & 8.3 & - \\ 
  2016 & 35 & 75 & - & - & - & - & - & - & - & - & 40 & - & - & - & 17 & - & - & 8.3 & - & 8.3 & 8.3 & 8.3 \\ 
    \midrule
\end{tabular}
}

\vspace{3mm}

\centerline{
\centering
\begin{tabular}{K{0.8cm}|K{0.2cm}Q{0.2cm}K{0.2cm}Q{0.2cm}K{0.2cm}Q{0.2cm}K{0.2cm}Q{0.2cm}K{0.2cm}Q{0.2cm}K{0.2cm}Q{0.2cm}K{0.2cm}Q{0.2cm}K{0.2cm}Q{0.2cm}K{0.2cm}Q{0.2cm}K{0.2cm}Q{0.2cm}K{0.2cm}Q{0.2cm}K{0.2cm}Q{0.2cm}K{0.2cm}Q{0.2cm}K{0.2cm}Q{0.2cm}K{0.2cm}Q{0.2cm}K{0.2cm}Q{0.2cm}K{0.2cm}Q{0.2cm}K{0.2cm}Q{0.2cm}K{0.2cm}Q{0.2cm}K{0.2cm}Q{0.2cm}}
\textbf{Dates} & \multicolumn{2}{c}{IWR} &  \multicolumn{2}{c}{IWF} &  \multicolumn{2}{c}{IWN} &  \multicolumn{2}{c}{IWM} &  \multicolumn{2}{c}{IYW} &  \multicolumn{2}{c}{IYR} &  \multicolumn{2}{c}{RSP} &  \multicolumn{2}{c}{EEM} &  \multicolumn{2}{c}{IWO} &  \multicolumn{2}{c}{IWV} \\
  \midrule
2003 & - & - & - & - & 8.3 & - & - & - & - & 8.3 & - & - & 8.3 & - & - & - & - & - & - & - \\ 
  2004 & - & - & - & 12 & - & - & - & - & - & 12 & - & - & 6.2 & - & - & - & - & - & - & - \\ 
  2005 & - & - & - & 8.3 & - & - & - & - & - & 8.3 & 30 & - & - & - & - & - & - & - & - & - \\ 
  2006 & - & - & 6.3 & - & - & - & 6.2 & - & - & 12 & - & - & - & - & - & - & - & - & - & - \\ 
  2007 & - & - & - & - & - & - & - & - & - & 8.3 & - & - & - & - & - & - & - & - & - & - \\ 
  2008 & - & - & - & - & - & - & - & - & - & - & - & - & - & - & - & - & - & - & - & - \\ 
  2009 & - & - & - & - & - & - & - & - & - & - & - & - & - & - & - & - & - & - & - & - \\ 
  2010 & - & - & - & - & - & - & - & - & 8.3 & 8.3 & - & - & - & - & - & - & - & - & - & - \\ 
  2011 & - & - & - & - & - & - & - & - & 8.3 & 8.3 & - & - & - & - & - & - & - & - & - & 8.3 \\ 
  2012 & - & - & - & 6.2 & - & - & - & - & 8.3 & - & - & - & - & - & - & - & - & - & - & 6.3 \\ 
  2013 & - & - & - & - & - & - & - & - & 8.3 & - & - & - & - & - & - & 6.3 & - & - & - & - \\ 
  2014 & 6.2 & - & - & - & - & - & - & - & - & - & - & - & - & - & - & 12 & - & 12 & - & - \\ 
  2015 & - & - & - & - & - & - & - & - & - & 6.2 & - & - & - & - & - & - & - & - & - & - \\ 
  2016 & - & - & - & - & - & - & - & - & - & - & - & - & - & - & - & - & - & - & - & - \\    
  \midrule
\end{tabular}
}

\caption{Sparse portfolio decisions (in percent) for DLMs (\ref{DLM1}) and (\ref{DLM2}) with $\delta_{F}= \delta_{\epsilon}= 0.97$ and $\delta_{c}= \delta_{\beta}= 0.9925$.  Shown are the selected portfolio decisions for the two targets: dense portfolio (left column in white) and SPY (right column in gray).  Note that annual weights are shown for brevity although portfolios are updated monthly.  In this dynamic portfolio selection, the regret threshold is $\kappa = 45\%$ for both targets.}
\label{DLMweightsk45}
\end{table}

We show the selected sparse decisions with the $\kappa=45\%$ threshold for the dense (portfolio optimized over all funds in white columns) and sparse (portfolio of only SPY in gray columns) target decisions in Table (\ref{DLMweightsk45}).  Each of these decisions possesses the property that, at each point in time, the satisfaction probability of investing in this decision versus the respective target is at least 45\%.  The portfolios are updated on a monthly basis; the table displays annual weights for brevity.  There are many decisions that will satisfy this threshold for each of the target (see for example the probability mass above 0.45 in Figure (\ref{pilamexample})).  In this case, we have added flexibility in which sparse decision to choose. We construct the sparse decisions so that at most one fund is selected or removed from month to month. For example, if the current portfolio at time $t$ has SPY, OEF, and IVV, two admissible portfolios at $t+1$ could include the funds be SPY, OEF, IVV, and EWG \textit{or} SPY and OEF assuming they both also satisfy the $\kappa$ threshold.

In Table (\ref{DLMweightsk45}), the sparse decision for the SPY target has larger allocations to SPY over the trading period compared with the sparse decision for the dense portfolio target.  Also, it possesses a consistent allocation to the US technology sector fund IYW.  In contrast, the sparse decision for the dense target often possesses a significant allocation to the Japanese equity specific fund EWJ.

\begin{figure}[H]
\renewcommand{\figurename}{Figure}
\centerline{
  \includegraphics[scale=.43]{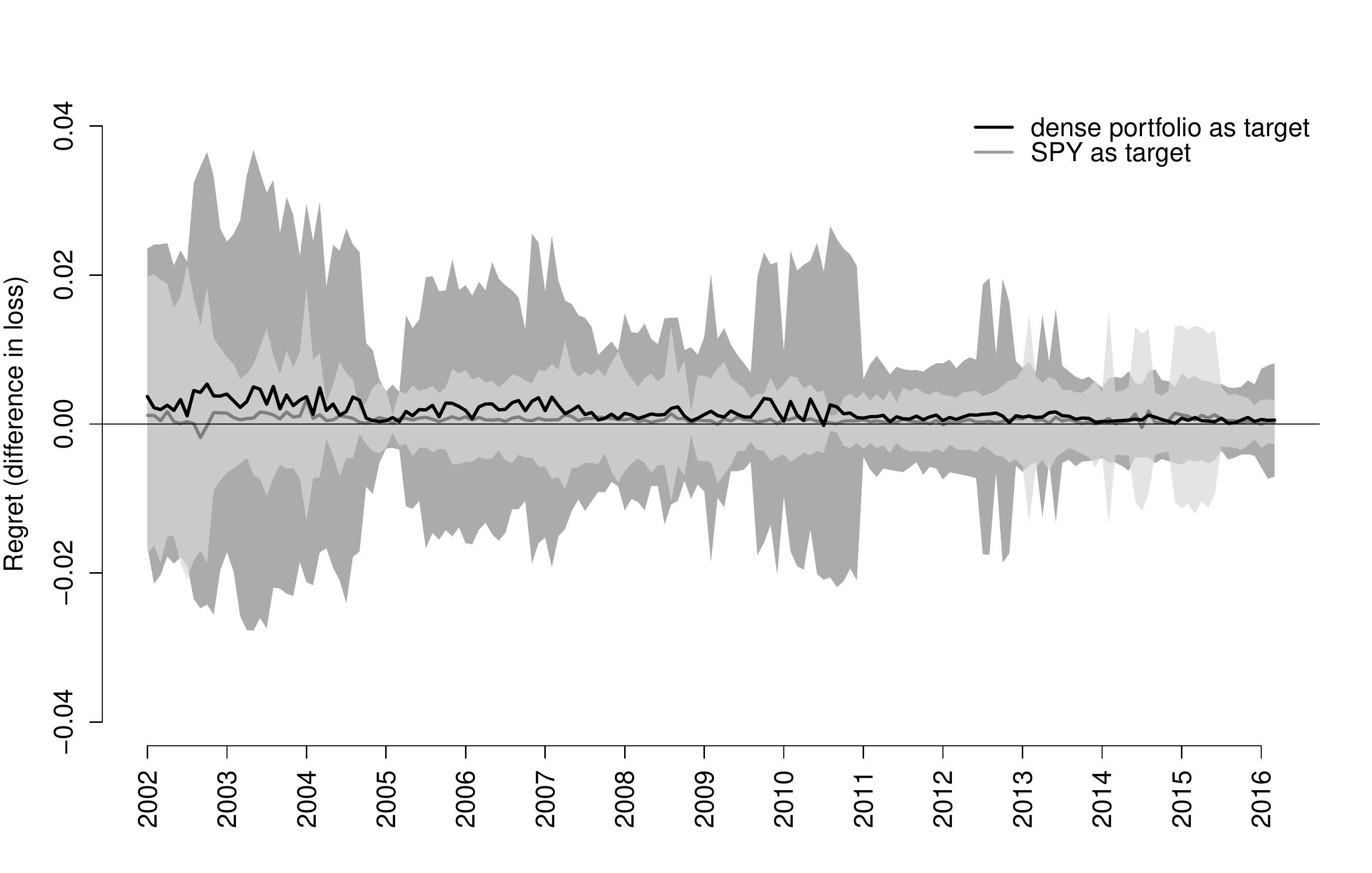}}
  \caption{The evolution of the ex ante regret distributions for the sparse decisions in Table (\ref{DLMweightsk45}) versus the two targets: dense portfolio (black) and SPY (gray).  The mean regret is shown by the lines and the surrounding areas represent the evolving centered 60\% posterior credible intervals.}
  \label{regretevolve1}
\end{figure}

Figure (\ref{regretevolve1}) displays the evolution of the regret distributions for the sparse decisions shown in Table (\ref{DLMweightsk45}).  The lines are the expected regret, and the surrounding areas correspond to the centered 60\% posterior credible intervals.  The expected regret for both decisions remains close to zero and for most investing periods is slightly above zero; this is by construction since we choose the sparse decision that satisfies the $\kappa=45\%$ threshold at each point in time.  Overall, these decisions do not result in much regret. Indeed, many of the enumerated long only decisions appear similar in terms of the next period log wealth utility.

The variance of the regret distributions in Figure (\ref{regretevolve1}) changes substantially over the investing period. The range of log cumulative wealth difference for the ``dense portfolio as target" at the beginning is large ($\sim0.98 \text{ to } 1.02$ on the cumulative wealth scale).  The sparse decision for the dense target collapses in variance around 2005 exactly when the sparse decision is very close to the dense portfolio. Notice also that the variance of regret for the sparse decision with SPY as the target is, in general, smaller than the dense target decision.  Since all of the enumerated decisions have at least 25\% of the portfolio allocated to SPY (the target itself) with other diversifying positions, it is intuitive that the uncertainty in regret should be smaller. 

The evolution of regret for the sparse decision with SPY as the target sheds light on another question: Are there diversification benefits of allocating to other funds in consideration?  Selecting among the 12,950 sparse decisions including up to four non-SPY funds, the expected regret appears to be essentially zero (see the gray line in Figure (\ref{regretevolve1})).  This analysis suggests that under the log cumulative wealth utility and considering the large set of enumerated decisions defined at the beginning of this section, the best sparse decision from an \textit{ex ante} perspective may be SPY itself!

The ex ante evolution of other metrics, such as the portfolio Sharpe ratio, may be studied for the sparse decisions displayed in Table (\ref{DLMweightsk45}).  The Sharpe ratio is not a utility since it is not a function of future returns $\tilde{R}_t$.  However, it is a function of our model parameters whose uncertainty is characterized by the posterior distribution.  Specifically, define the predictive portfolio Sharpe ratio:
	\begin{equation} \label{regretdefnSR}
	\begin{split}
		\mathcal{SR}(w_t,\Theta_t) &= w_t^{T}\mu_t / (w_t^{T}\Sigma_t w_t)^{1/2},
		\\
		\rho_{\mathcal{SR}}(w_{\lambda_t}^{*},w_t^*,\Theta_t) &= \mathcal{SR}(w_t^*,\Theta_t)-\mathcal{SR}(w_{\lambda_t}^{*},\Theta_t),
	\end{split}
\end{equation}where $\rho_{\mathcal{SR}}(\cdot)$ is predictive in the sense that future returns $\tilde{R}_t$ conditional on the model parameters are integrated out. This portfolio metric differs from the Kelly criterion loss in that it focuses on a ratio of the portfolio expected return and variance. It may be used as an exploratory tool to accompany selection from regret-based portfolio selection.

\begin{figure}[H]
\renewcommand{\figurename}{Figure}
\centerline{
  \includegraphics[scale=.43]{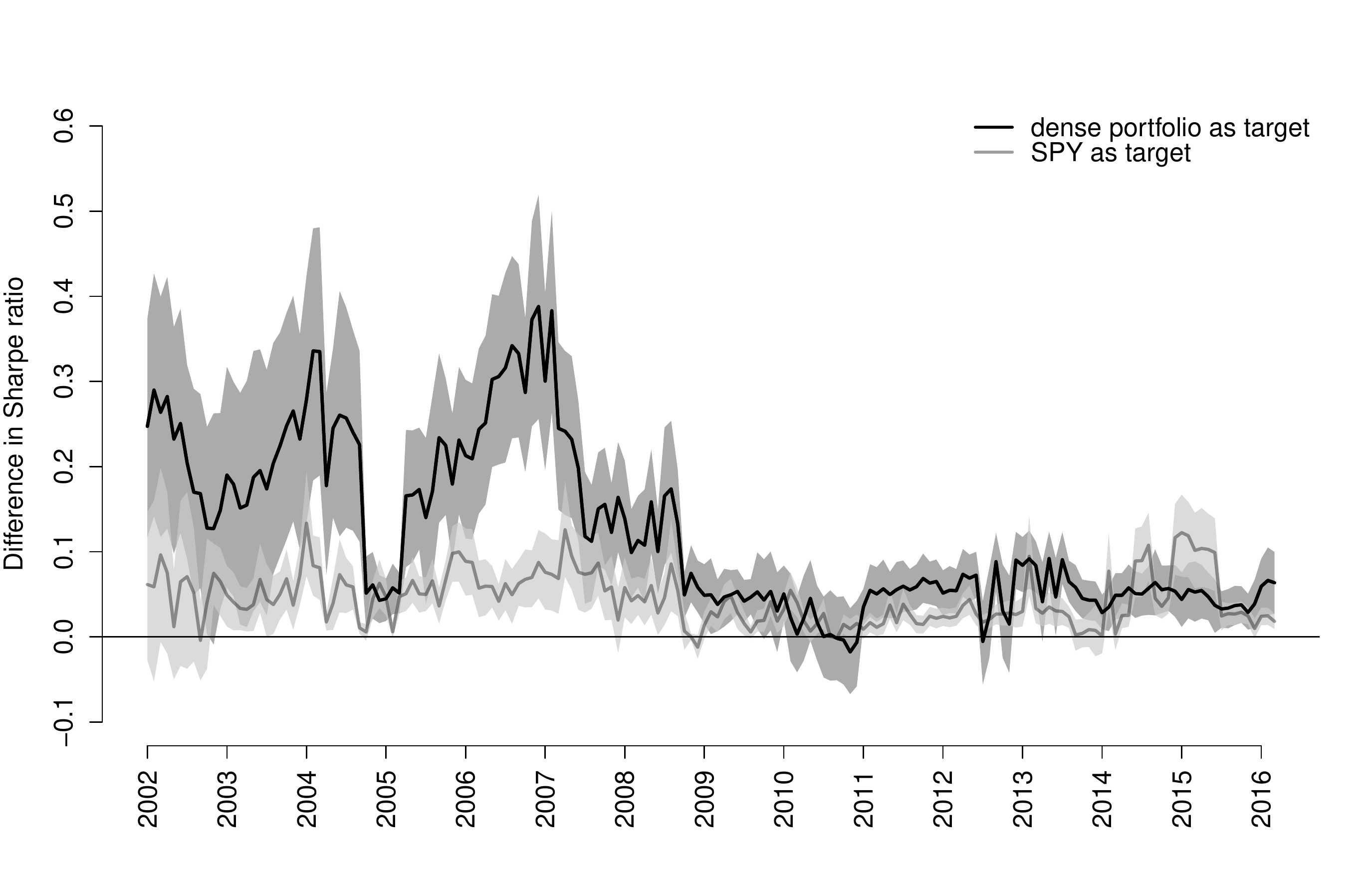}}
  \caption{The evolution of the ex ante ``Difference in annualized Sharpe ratio" distributions for the sparse decisions in Table (\ref{DLMweightsk45}) versus the two targets: dense portfolio (black) and SPY (gray).  The mean regret is shown by the lines and the surrounding areas represent the evolving centered 60\% posterior credible intervals.}
  \label{SRevolve1}
\end{figure}

We utilize this ``difference in Sharpe ratio" distribution in an exploratory fashion in Figure (\ref{SRevolve1}) shown on an annualized scale.  The evolution of the difference in Sharpe ratio is similar to the regret in Figure (\ref{regretevolve1}). In this case, a larger positive difference in Sharpe ratio means the selected sparse decision possesses a smaller return-risk tradeoff compared to the target decision.  The sparse decision for the dense target is larger variance and trends around larger positive values compared with the sparse decision for the SPY target.  The rationale for these features is similar: The enumerated sparse decisions are constructed to contain SPY, so the Sharpe ratios (like the loss) of the sparse decision and the SPY target decision will often be close. Following the financial crisis around 2009, the difference in Sharpe ratio stabilizes at lower values for both sparse decisions.

\subsubsection{What happens when $\kappa$ is varied?}

\begin{figure}[H]
\renewcommand{\figurename}{Figure}
\centerline{
  \includegraphics[scale=.4]{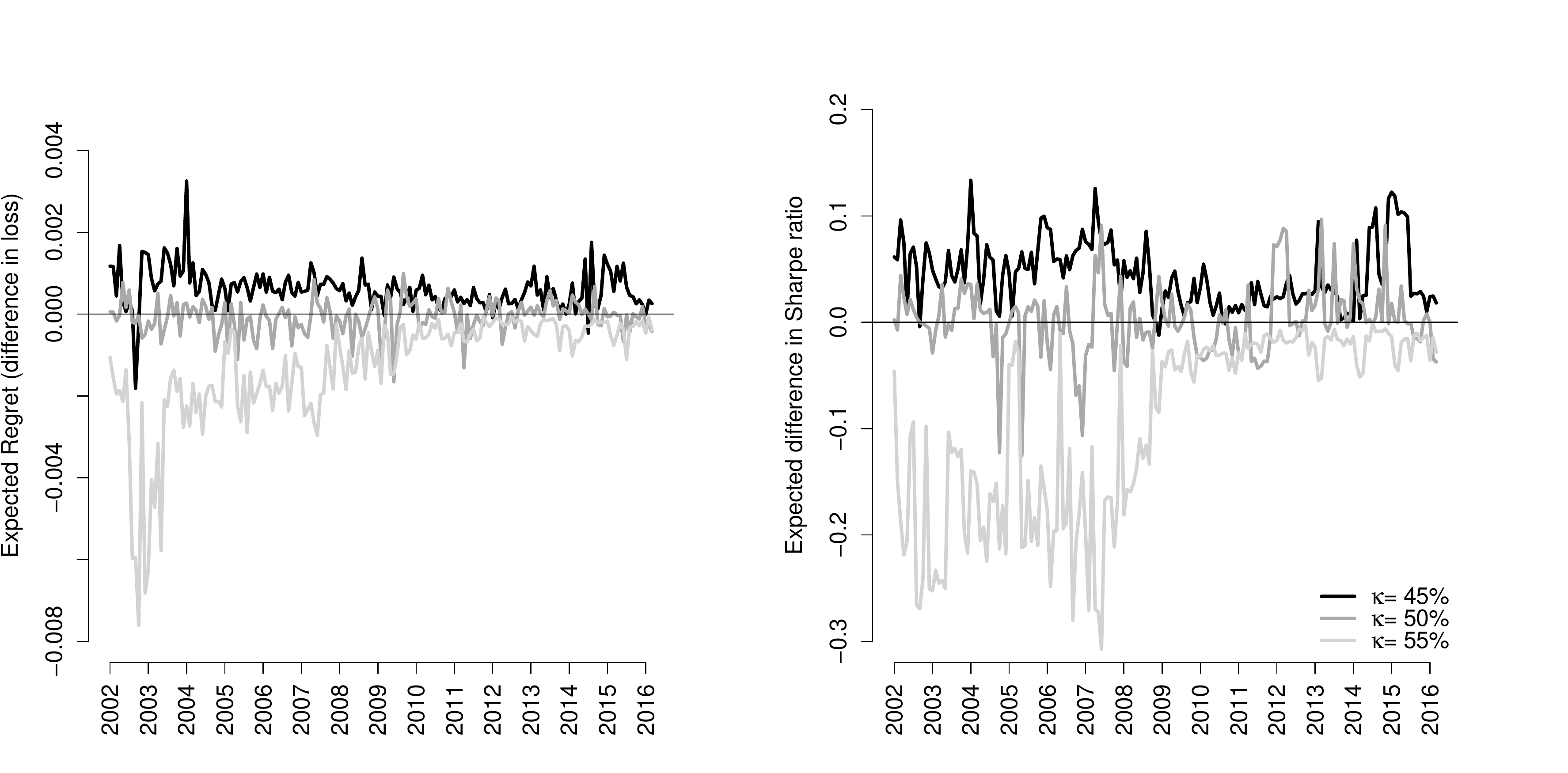}}
  \caption{Expected regret (left) and expected difference in Sharpe ratio (right) for sparse decisions with SPY as the target.  These metrics are shown for three regret thresholds (the lower bound on probability of satisfaction): $\kappa=45\%$ (black), $50\%$ (dark gray), and $55\%$ (light gray).  Note that as the lower bound on the probability of satisfaction increases, both the expected regret and difference in Sharpe ratio tend to decrease for the selected sparse decisions.}
  \label{expectedregret}
\end{figure}

The selection of dynamic portfolio decisions will change based on the regret threshold $\kappa$.  In Figure (\ref{expectedregret}), we show how expected regret and difference in Sharpe ratio (on an annualized scale) change for selected sparse decisions using the SPY target.  The evolution of these metrics is shown for sparse decisions constructed using three $\kappa$ thresholds: $\kappa=45\%$ (black), $50\%$ (dark gray), and $55\%$ (light gray).  The black lines in both figures correspond to the ``SPY as target" paths in Figures (\ref{regretevolve1}) and (\ref{SRevolve1}), now compared to other $\kappa$ choices. 

Since $\kappa$ is a lower bound on the satisfaction probability, increasing this lower bound should lead to dynamic sparse decisions with generally smaller regret.  In other words, if the investor would like to be satisfied with higher probability, a ``lower regret"-sequence of sparse decisions should be selected. Figure (\ref{expectedregret}) demonstrates this when SPY is the target.  Larger $\kappa$ generally lead to smaller expected regret and difference in Sharpe ratio paths.  The $\kappa=55\%$ sparse decision leads to expected regret and difference in Sharpe ratio that are mostly negative from 2002 through 2016, indicating that portfolios with SPY and diversifying funds may be preferred to just SPY alone at this high satisfaction threshold.  However, these differences in expectation are still close to zero and small, especially for the evolution of expected regret. 

\subsubsection{Enumerated decisions without using the utility}

The enumerated decisions considered up to this point are constructed by optimizing the integrated approximate loss.  An investor might prefer to construct decisions without any consideration for utility and statistical model. An equal-weighted portfolio (where each of the $N$ assets is given weight $1/N$) is one such example of a ``utility agnostic" decision.  Financial practitioners often advocate for this decision because of its  out of sample performance and purity in not involving errors in the statistical model and optimization procedure \citep{demiguel2007optimal}.  The regret-based procedure can readily accommodate a set of decisions with these characteristics as well.

In the following analysis we consider the set of sparse enumerated equal-weighted portfolios with up to four funds. This amounts to $\Sigma_{i=1}^4 {25 \choose i} = 15,275$ decisions to choose among at each point in time. We choose the ``dense $1/N$" portfolio as the target decision.  This target has 4\% invested in each of the 25 funds. To remain consistent with the previous analysis, we consider selection when $\kappa=45\%$.

\hspace{3mm}
\begin{table}[H]

\footnotesize
\centerline{
\begin{tabular}{K{0.8cm}|K{0.4cm}K{0.4cm}K{0.4cm}K{0.4cm}K{0.4cm}K{0.4cm}K{0.4cm}K{0.4cm}K{0.4cm}K{0.4cm}K{0.4cm}K{0.4cm}K{0.4cm}K{0.4cm}K{0.4cm}K{0.4cm}K{0.4cm}K{0.4cm}K{0.4cm}K{0.4cm}K{0.4cm}K{0.4cm}K{0.4cm}K{0.4cm}K{0.4cm}K{0.4cm}K{0.4cm}K{0.4cm}K{0.4cm}K{0.4cm}K{0.4cm}K{0.4cm}K{0.4cm}K{0.4cm}K{0.4cm}K{0.4cm}K{0.4cm}K{0.4cm}K{0.4cm}K{0.4cm}}
\textbf{Dates} & SPY & DIA & EZU & EWU & EWY & EWJ & OEF & EEM & IVE & EFA & IWP & IWR & IWF & IWO & IWM & IYW & XLI \\
  \midrule
2003 & - & - & - & - & - & - & - & - & - & 25 & 25 & - & - & - & 25 & 25 & - \\ 
  2004 & - & - & - & - & 33 & 33 & - & - & - & - & - & - & - & - & - & 33 & - \\ 
  2005 & - & - & - & - & 25 & - & 25 & - & - & - & - & - & - & - & - & 25 & 25 \\ 
  2006 & - & - & - & 33 & - & - & - & - & - & - & - & 33 & - & - & - & 33 & - \\ 
  2007 & - & - & - & 33 & - & - & - & - & - & - & 33 & - & 33 & - & - & - & - \\ 
  2008 & - & - & 25 & - & - & - & 25 & - & - & - & 25 & - & 25 & - & - & - & - \\ 
  2009 & - & 33 & - & - & - & 33 & - & - & - & - & - & - & - & - & - & 33 & - \\ 
  2010 & - & - & - & - & - & 33 & - & 33 & - & - & - & - & - & - & - & 33 & - \\ 
  2011 & - & - & - & - & - & - & - & 33 & - & 33 & 33 & - & - & - & - & - & - \\ 
  2012 & - & - & - & 33 & 33 & - & 33 & - & - & - & - & - & - & - & - & - & - \\ 
  2013 & - & - & - & - & - & - & 33 & - & 33 & - & - & - & - & 33 & - & - & - \\ 
  2014 & 33 & - & - & - & - & - & - & - & - & 33 & - & - & - & 33 & - & - & - \\ 
  2015 & - & - & 25 & - & - & 25 & - & - & 25 & - & - & - & - & - & 25 & - & - \\ 
  2016 & - & - & - & 33 & - & 33 & - & - & 33 & - & - & - & - & - & - & - & - \\  
  \midrule
\end{tabular}
}

\caption{Sparse portfolio decision (in rounded percent) for DLMs (\ref{DLM1}) and (\ref{DLM2}) with $\delta_{F}= \delta_{\epsilon}= 0.97$ and $\delta_{c}= \delta_{\beta}= 0.9925$.  Each point in time represents an equal-weighted portfolio and corresponding $\lambda_t$ such that the decision satisfies the $\kappa=45\%$ threshold.  The target decision is the equal-weighted portfolio of all 25 funds -- also known as the dense $1/N$ portfolio.  Note that annual weights are shown for brevity although portfolios are updated monthly.} 
\label{DLMweightsk45EW}
\end{table}

The weights for the selected portfolio decision are shown in Table (\ref{DLMweightsk45EW}).  At the $\kappa=45\%$ threshold, all portfolios have either three or four funds included, and the portfolio decisions have sustained exposures to EWJ (Japanese equity) and IYW (technology) throughout the investing period.  

This approach to equal-weighted portfolio selection possesses innovative and important features that should be highlighted.  While traditional ``$1/N$" approaches avoid the investor's utility and model for future asset returns altogether, the regret-based procedure still accounts for both.  The decisions themselves may be constructed without a utility or model in mind, but the characterization of regret must involve the utility and model.  Regret is computed by the difference in \textit{utility} between the target and sparse decisions, and it is a random variable that may be simulated using the \textit{model} for future asset returns.  Regardless of the set of decisions considered by the investor, the utility and model will always play a crucial role in a regret-based selection procedure.

\subsubsection{Ex post decision analysis}

In this section, we consider the realized performance of the three sparse portfolio decisions presented in Tables (\ref{DLMweightsk45}) and (\ref{DLMweightsk45EW}) relative to their target decisions. We present out of sample statistics for the six decisions in Table (\ref{OOStablenew}). Shown is the annualized Sharpe ratio, standard deviation of return, and mean return.

The sparse enumerated decision for the dense target performs similarly to the dense target.  This is comforting -- this sparse decision is a dynamic portfolio that is allocated to the market (SPY) and at most four other diversifying positions, and its out of sample performance is comparable to the dense portfolio optimized over all 25 funds at each time.

\begin{table}[H]
\begin{center}
\footnotesize

\begin{tabular}{l*3{>{\centering\arraybackslash}m{0.4in}} @{}m{0pt}@{}}
 & \multicolumn{3}{c}{out-of-sample statistics} \\[1mm]
 \cline{2-4}
 &  Sharpe ratio & s.d. & mean return\\[.1ex] 
  \hline
sparse enumerated - dense as target & 0.40 & 14.98 & 6.02  \\ 
dense & 0.45 & 14.41 & 6.47  \\ 
  \hline
    sparse enumerated - SPY as target & 0.43 & 14.65 & 6.28 \\ 
    SPY & 0.43 & 14.63 & 6.28  \\ 
  \hline
  sparse EW enumerated - dense $1/N$ as target & 0.49 & 16.71 & 8.15  \\ 
 dense $1/N$ & 0.44 & 16.47 & 7.32  \\ 
  \hline
\end{tabular}

\end{center}
\caption{Comparison of out of sample statistics for the six portfolio strategies considered over the investing period February 2002 to May 2016. All statistics are presented on an annualized scale. ``EW" refers to the equal-weighted portfolio decision.}
\label{OOStablenew}
\end{table}

The sparse enumerated decision for the SPY target is equally interesting.  Since the SPY target is a sparse decision itself, comparison of it with selected sparse decisions provides insight into whether or not one should diversify away from investing in just the \textit{Market}. The out of sample performance shown in rows three and four of Table (\ref{OOStablenew}) display similar performance of the sparse enumerated decision and SPY.  Even after considering 12,950 sparse decisions containing up to four funds other than SPY -- the diversification benefits of exposure beyond SPY are negligible. The decisions that are \textit{ex ante} better than SPY with $45\%$ probability turn out to help out little \textit{ex post}.  Note that this conclusion is with respect to the next period cumulative wealth utility.  Future work will involve consideration of other utilities and compare how their selection and ex ante/post analyses differ.

While the sparse optimal strategies both underperform their targets, the sparse equal-weighted strategy slightly outperforms its dense $1/N$ target.  This is shown in rows five and six of Table (\ref{OOStablenew}). Interestingly, its out of sample performance even exceeds its sparse optimal counterparts shown in rows one and three in the Table.

\section{Discussion}

This paper presents a new approach to portfolio selection based on an investor-specified regret tolerance.  A loss function defined by the expected portfolio growth rate is used in tandem with a new procedure that separates statistical inference from selection of a sparse dynamic portfolio.  We illustrate the procedure using a set of exchange-traded funds.  After analyzing two target decisions: (\textit{i}) A dense portfolio of all available assets, and (\textit{ii}) A portfolio comprised of a single market fund, we find that selected sparse decisions differ little from their targets in terms of utility; especially after taking into account uncertainty. This finding persists ex post, and a variety of sparse decisions perform similarly to their target decisions on a risk adjusted return (or Sharpe ratio) basis.

The procedure offers a fresh approach to portfolio selection.  While traditional approaches typically focus on either the careful modeling of parameters or the optimization procedure used to calculate portfolio weights, regret-based selection combines both through analysis of the regret random variable.  Portfolio decisions that are \textit{parsimonious} in nature are then evaluated in a framework that incorporates \textit{uncertainty} in the investor's utility. 

Areas of future research include alternative utility specifications.  Two relevant examples are: (\textit{i}) incorporation of fees and (\textit{ii}) minimization of transaction costs.  In each case, a variant of Loss function (\ref{lossorig}) may be considered.  Fees of the funds can be incorporated directly into the vector of future returns.  For example, suppose a vector of expense ratios (percentage fee charged of total assets managed) of all funds were given by $\tau$.  The vector of future returns within the loss function may be adjusted by $\tau$ to reflect an investor's sensitivity to fees:
\begin{equation}
	\begin{split} \label{optprobconclusion}
		\mathcal{L}(w_t,\tilde{R}_t) &= - \log \left( 1 + \sum_{k=1}^{N} w_t^k(\tilde{R}_t^k - \tau_{k}) \right) ,
	\end{split}
\end{equation}where $\tilde{R}_t^k - \tau_{k}$ is the net return on investment in for fund $k$. 

Sensitivity to transaction costs can be similarly accounted for by modifying the complexity (penalty) function $\Phi$.  This can be accomplished by penalizing the difference in consecutive weight vectors through time, $w_{t} - w_{t-1}$.  An example penalty function would look like:
\begin{equation}
	\begin{split} \label{optprobconclusion2}
	\Phi(\lambda_t^1,\lambda_t^2,w_t) &=  \lambda_t^1 \norm{ w_t }_{1} + \lambda_t^2 \norm{ w_t - w_{t-1} }_{1}.	
\end{split}
\end{equation}This penalty is designed to encourage sparsity as well as slow movement in portfolio positions over time so as to avoid frequent buying and selling of assets.  It poses an interesting challenge since there are two penalty parameters ($\lambda_t^1 \text{ and } \lambda_t^2$) that must be chosen.  This is precisely where the regret-based framework has merit. Portfolio decisions indexed by these two penalties can be mapped to a digestible single probability of regret.  Then, selection of an appropriate $\{\lambda_t^1,\lambda_t^2\}$ pair can be done in this intuitive ``regret space".

The remainder of this section takes a step back and discusses the modularity and important features of regret-based portfolio selection. The methodology is intended to be general -- the particular loss, model and dataset used for the empirical analysis are only chosen to demonstrate how the procedure works in practice.  The primitive components are: (\textit{i}) a utility function characterizing investor preferences, (\textit{ii}) a complexity function measuring how ``simple" a portfolio decision is, (\textit{iii}) a statistical model, and (\textit{iv}) the investor's regret tolerance; where regret is defined as a difference in utility.  The regret tolerance stitches together the first two primitives by answering the question: How does the investor view the \textit{tradeoff} between her utility and portfolio complexity?  Using the utility and posterior distribution defined by the statistical model (primitive three), one can construct a mapping between a set of penalty parameters $\{ \lambda_t \}$ and probabilities of regret (as displayed by the right vertical axis in Figure (\ref{deltaexample})).  However, this is not enough.  The collection of portfolio decisions indexed by $\lambda_t$ must be distilled down to one.  The fourth primitive accomplishes this by placing an upper bound on the probability of regret; a portfolio that satisfies this upper bound is selected.  By incorporating the four primitives, the main (and surprising) feature of this methodology is that a \textit{static} regret threshold produces a sequence of \textit{dynamic} portfolio decisions, one for each investing period.

\appendix

\section{Penalized optimization} \label{applossfunopt}
In this section, we show how to minimize a penalized loss function of the following form:
\begin{equation}
	\begin{split}
		\textbf{L}_{\lambda_t}(w_t) =  \frac{1}{2}w_t^{T}\overline{\Sigma}_t w_t - w_t^{T}\overline{\mu}_t + \lambda_t \norm{ w_t }_{1}.
	\end{split}
\end{equation}

 Completing the square in $w_t$, dropping terms that do not involve the portfolio weights, and defining $\overline{\Sigma}_t = L_tL_t^{T}$, we rewrite the first two terms of the loss function as an $l_{2}$-norm:
 \begin{equation}
	\begin{split}
	\textbf{L}_{\lambda_t}(w_t) &=\frac{1}{2}w_t^{T}\overline{\Sigma}_t w_t - w_t^{T}\overline{\mu}_t + \lambda_t \norm{ w_t }_{1}
	 \\
	 &\propto \frac{1}{2}(w_t-\overline{\Sigma}_t^{-1}\overline{\mu}_t)^{T}\overline{\Sigma}_t(w_t-\overline{\Sigma}_t^{-1}\overline{\mu}_t) + \lambda_t \norm{ w_t }_{1}
	 \\
	 &= \frac{1}{2}(w_t-L_t^{-T}L_t^{-1}\overline{\mu}_t)^{T}L_tL_t^{T}(w_t-L_t^{-T}L_t^{-1}\overline{\mu}_t) + \lambda_t \norm{ w_t }_{1}
	 \\
	 &= \frac{1}{2} \left[ L_t^{T}(w_t-L_t^{-T}L_t^{-1}\overline{\mu}_t) \right]^{T} \left[ L_t^{T}(w_t-L_t^{-T}L_t^{-1}\overline{\mu}_t) \right] + \lambda_t \norm{ w_t }_{1}
	 \\
	 &= \frac{1}{2} \norm{ L_t^{T}w_t-L_t^{-1}\overline{\mu}_t }_{2}^{2} + \lambda_t \norm{ w_t }_{1}.
	\end{split}
\end{equation} The optimization of interest is written as:
\begin{equation}
	\begin{split} \label{optprobfinal}
		 \min_{w_t \geq 0} \hspace{3mm} \frac{1}{2} \norm{ L_t^{T}w_t-L_t^{-1}\overline{\mu}_t }_{2}^{2} + \lambda_t \norm{ w_t }_{1}.
	\end{split}
\end{equation}
Optimization (\ref{optprobfinal}) is now in the form of standard sparse regression loss functions, \citep{Tib}, with covariates, $L_t^{T}$, data, $L_t^{-1}\overline{\mu}_t$, and regression coefficients, $w_t$.  We may optimize (\ref{optprobfinal}) conveniently using existing software, such as the \texttt{glmnet} package of \cite{friedman2010regularization}.

\newpage
\bibliographystyle{imsart-nameyear.bst}
\bibliography{DSSPortfolioOptimization.bib}

\end{document}